\documentclass[%
 aip,
 jcp,
 amsmath,amssymb,
reprint,
]{revtex4-1}

\usepackage{hyperref}
\usepackage{cleveref}
\usepackage{color}
\usepackage{graphicx}% Include figure files
\usepackage{dcolumn}% Align table columns on decimal point
\usepackage{bm}% bold math
\DeclareMathOperator{\tr}{Tr}
\usepackage{epstopdf}

\begin{document}

%\preprint{}
%AIP/123-QED}

\title{Quantum dynamics of the intramolecular vibrational energy redistribution in OCS: From localization to quasi-thermalization}

\author{J. B. P\'erez}
\author{J. C. Arce}
\email{julio.arce@correounivalle.edu.co}
\affiliation{Departamento de Qu\'imica, Universidad del Valle, A.A. 25360, Cali, Colombia}

\date{\today}

{
\begin{abstract}

We report a fully quantum-dynamical study of the intramolecular vibrational energy redistribution (IVR) in the electronic ground state of carbonyl sulfide (OCS), which is a prototype of an isolated many-body quantum system with strong internal couplings and non-Rice-Ramsperger-Kassel-Marcus (RRKM) behavior.
We pay particular attention to the role of many-body localization and the approach to thermalization, which currently are topics of considerable interest, as they pertain to the very foundations of statistical mechanics and thermodynamics.
We employ local-mode (valence) coordinates and consider initial excitations localized in one local mode, with energies ranging from low to near the dissociation threshold, where the classical dynamics have been shown to be chaotic.
We propagate the nuclear wavepacket on the potential energy surface by means of the numerically exact multiconfiguration time-dependent Hartree method and employ mean local energies, time-dependent and time-averaged populations in quantum number space, energy distributions, entanglement entropies, local population distributions, microcanonical averages, and dissociation probabilities, as diagnostic tools.
This allows us to identify a continuous localization $\rightarrow$ delocalization transition in the energy flow, associated with the onset of quantum chaos, as the excitation energy increases up to near the dissociation threshold.
Moreover, we find that at this energy and $\sim$1 ps the molecule nearly thermalizes.
Furthermore, we observe that IVR is so slow that the molecule begins to dissociate well before such quasi-thermalization is complete, in accordance with earlier classical-mechanical predictions of non-RRKM behavior.

\end{abstract}
}

%\pacs{31.10.+z, 31.15.V-, 31.50.-x}
%PACS, the Physics and Astronomy Classification Scheme.


\maketitle

\section{INTRODUCTION}

%The Hamiltonian of a composite quantum system composed of a system coupled to a bath is given by
%
%\begin{equation}
%\hat{H}=\hat{H}_{S}+\hat{H}_{B}+\hat{H}_{I}
%\end{equation}
%
%where $\hat{H}_{S}$ is the system Hamiltonian, $\hat{H}_{B}$ is the bath Hamiltonian, and $\hat{H}_{I}$ is the operator representing the interaction between them. 
%The thermal equilibrium state of the system is represented by a reduced density operator, which is given by the trace over the bath states of the density operator of the composite system. In the canonical ensemble, for instance, the reduced density operator is \cite{Huse}
%
%\begin{equation}
%\hat{\rho}_{s}=\textrm{Tr}_{B} \left[ %\frac{e^{-\hat{H}/k_{B}T}}{Z} \right],
%\end{equation}
%
%where $Z$ is the canonical partition function of the composite system. In the limit where the system-bath interaction is negligible, the reduced density operator  reduces to the Boltzmann form
%
%\begin{equation}
%\hat{\rho}_{s}=\frac{e^{-\hat{H}_{S}/k_{B}T}}{Z_{s}},
%\end{equation}
%
%where now $Z_s$ is the canonical partition function of the system alone.

The questions of when and how an isolated many-body quantum system reaches a state that can be characterized as a thermal equilibrium state, a process called thermalization, lies at the foundation of statistical mechanics and thermodynamics.\cite{Huse,Gogolin,Dalessio}
Only until very recently these questions have begun to be satisfactorily answered, thanks to developments in experimental methods for the control of quantum systems with many degrees of freedom (DOFs), computing power, numerical techniques, and mathematical methods.\cite{Gogolin}

In this paper, we regard thermalization a la von Neumann,\cite{Neumann} i.e., we assume that the system is initially prepared in a nonstationary state and subsequently expectation values of observables that do not commute with the Hamiltonian relax to the value predicted by the microcanonical distribution.\cite{Dalessio}
Moreover, we consider a composite system in a pure state evolving unitarily under the time-dependent Schr\"odinger equation (TDSE). Hence, each subsystem thermalizes by virtue of its interactions with the other subsystems, which can be considered as composing a bath.\cite{Gogolin} 

%The process by which an isolated composite quantum system reaches a statistical thermal equilibrium state, called thermalization, results from the entanglement between all the subsystems that compose it.\cite{Huse} This phenomenon is usually supposed to occur in the infinite-size and infinite-time limits. However, it has been shown that it can emerge in finite systems and in finite time, for example in a six-site Bose-Hubbard chain model,\cite{Kaufman} and in a system composed of three interacting superconducting qubits.\cite{Neill}

%An excited molecule interacting with the surroundings can thermalize through vibrational relaxation \cite{Tramer}. Furthermore, if a molecule is large enough, each vibrational degree of freedom (DOF) can be considered as an open quantum system interacting with an environment composed by the other vibrational DOFs.

A molecule in a definite electronic state can be considered as a composite system, with the vibrational modes playing the role of the subsystems. Due to the anharmonicity of the potential energy surface (PES), it is impossible to find a coordinate system that allows the definition of exactly uncoupled vibrational modes. Therefore, if one of the modes is initially excited, a fraction of the excitation energy will inexorably flow into the other modes. Such process is termed intramolecular vibrational energy redistribution (IVR).\cite{Uzer,Field} Hence, according to the previous paragraph, an isolated molecule could thermalize through IVR.
%, provided this process be ergodic.
In fact, the hypothesis that IVR leads to thermalization before any bond is broken is basic in the statistical Rice-Ramsperger-Kassel-Marcus (RRKM) theory of unimolecular reactions.\cite{Marcus,Baer-Hase} The success of this theory\cite{Baer-Hase} provides strong evidence that many isolated molecules indeed can rapidly thermalize.

A key feature of thermalization is its independence of the initial state of the system, i.e., that the subsystems do not retain memory of their initial conditions.\cite{Huse,Gogolin} In the contexts of IVR and RRKM theory, the violation of this condition is called mode specificity.\cite{Baer-Hase}
The failure of a system to thermalize can be considered as the signature of the so-called many-body localization (MBL), which is the extension of Anderson localization to interacting composite systems.\cite{Huse,Gogolin}
%The quantum phase transition that takes place between the thermal and the MBL phases, as parameters in the Hamiltonian are varied, has been characterized in several systems.\cite{Huse}
Logan and Wolynes\cite{Wolynes1} have characterized the transition from weak to global IVR in a manner analogous to the characterization of the transition between the MBL and the thermal phases.\cite{Huse}
%The hypothesis that the dynamics of the vibrational degrees of freedom (DOFs) of molecules are ergodic is basic in the statistical Rice-Ramsperger-Kassel-Marcus (RRKM) theory of unimolecular reactions.\cite{Marcus} This theory further assumes that IVR is fast enough to lead to internal thermalization of the molecule before any bond is broken. 
Therefore, some of the observed deviations from the predictions of RRKM theory\cite{Bunker73,Marcus84,Rice90,Armenise92,Tonner2000,Utz2009} can be attributed to MBL. (Even when energy does flow, deviations can still appear if such flow is slower than the bond breaking.\cite{Gruebele2})

Strictly speaking, a system thermalizes only in the thermodynamic limit, i.e., when it possesses an infinite number of DOFs and evolves for an infinite time. However, it has been shown that, for practical purposes, thermalization can emerge in relatively small quantum systems in finite time, for example in a Bose-Einstein condensate of $^{87}$Rb atoms loaded into an optical lattice, which can be modeled as a six-site Bose-Hubbard chain,\cite{Kaufman} and in a model of three interacting superconducting qubits.\cite{Neill}
The systems that have recently been employed to address experimentally the issue of thermalization are of optical and/or condensed phase nature, taking advantage of sophisticated techniques for their near isolation.\cite{Gogolin,Kaufman,Neill} Moreover, commonly their relaxation to equilibrium is studied after applying a quench, i.e., a sudden change in the Hamiltonian.\cite{Kaufman,Santos}
On the other hand, for a long time chemists have been studying IVR and unimolecular decomposition reactions in the gas phase at low pressure, where the molecules are well isolated.\cite{Uzer,Field,Baer-Hase} In this case, the Hamiltonian cannot be manipulated, but, instead of applying quenches, specific initial nonstationary states can be prepared employing optical selective excitation\cite{Baer-Hase} and control\cite{Brumershapiro2} schemes. Besides, since the variety of molecules is virtually unlimited, wide ranges of couplings, densities of states (DOSs), and sizes are available. Hence, motivated by the observation that molecules can also serve as convenient laboratories for the study of fundamental aspects of statistical mechanics and thermodynamics, in this work we examine the intramolecular quantum dynamics in the smallest system that can exhibit IVR, namely the triatomic molecule, which possesses only three internal DOFs. We aim at revealing the quantal mechanism of the process and the energy and time scales involved, which permits us to address the issues of thermalization, mode specificity, and localization.

%An excited molecule interacting with the surroundings can thermalize through vibrational relaxation.\cite{Tramer} Furthermore, if a molecule is large enough, each vibrational degree of freedom (DOF) can be considered as an open quantum system interacting with an environment composed by the other vibrational DOFs.
%Either for calculation or analysis purposes, in quantum-mechanical studies of IVR a zero-order Hamiltonian is employed, whose eigenstates are used for representing the vibrational wavefunction of the molecule.
%Since these are not the true vibrational eigenstates, they are coupled. It is convenient to employ a representation where such coupling is as small as possible. At low excitation energy, the coupling between normal modes is usually weak.

In a quantum-mechanical description of IVR, the vibrational modes are represented by suitable single-variable zero-order Hamiltonians, whose eigenstates can be used as a single-mode basis for representing the multi-mode vibrational wavefunction of the molecule.
Under the assumption that the basis states form a quasi-continuum, the process has been described by the Fermi Golden Rule.\cite{Jortner} However, experimental studies have revealed that the rate of IVR does not always increase with the DOS in the manner predicted by this rule.\cite{McDonald1} To account for this observation, the process has been treated as a population flow from the initial state into the tier structure formed by the zero-order states.\cite{sibert,Stuchebrukhov1,Stuchebrukhov2} More recently, IVR has been described as a population diffusion in the so-called quantum number space (QNS) defined by the zero-order states.\cite{Gruebele2,Gruebele1,leitner} By varying the parameters of the Hamiltonian, those approaches have provided valuable information about general features of the process.
Nevertheless, for a realistic and detailed description of the intramolecular dynamics in a specific system, one must employ the full (Born-Oppenheimer) vibrational Hamiltonian with an accurate PES. Since, for polyatomic molecules, this Hamiltonian possesses a large and complicated discrete eigenspectrum, together with a continuous eigenspectrum, exact diagonalization\cite{Gogolin} is extremely demanding. Therefore, the time evolution must be generated by means of a numerical method.

As specific system, we chose the carbonyl sulfide (OCS) molecule, because it has long served as a prototype for the investigation of IVR and thermalization. \cite{CarterBrumer, *carter83,DavisWagner,davis84,Davis,Martens,Chandre,PaskaukasPRL,paskaukas,Gibson}
Carter and Brumer\cite{CarterBrumer,*carter83} performed a classical-mechanical study of a nonrotating model, finding a regular (quasi-periodic) behavior at energies below $\sim 65\%$ of the dissociation threshold (which corresponds to the breaking of the C-S bond) and irregular (chaotic) behavior above this energy. Moreover, they found that at $\sim 90\%$ of the dissociation energy, despite this chaotic behavior, after 2.4 ps local-mode energies did not relax to the average values predicted by the microcanonical ensemble.
Davis and Wagner\cite{DavisWagner} extended the study of Carter and Brumer to a model with the molecule confined to a plane and including energies above the dissociation threshold, finding incomplete IVR and mode-specificity even after 45 ps.
Further investigations focused attention on finding and characterizing the phase space bottlenecks responsible for the nonstatistical IVR in collinear\cite{davis84,Davis} and planar\cite{Martens,Chandre,PaskaukasPRL,paskaukas} models.
Gibson and coworkers\cite{Gibson} compared the classical and quantal dynamics for the collinear model by means of the Husimi distribution, concluding that the bottlenecks indeed constrain wave packets to evolve within one (or a combination of) chaotic region(s), although the trapping in the quantal case can be much stronger than in the classical one.

%All the abovementioned studies employ an empirical PES. Thus, it is highly desirable to determine to what extent their observations remain valid when a more accurate PES is used. Moreover,

To date, the full three-DOF quantum dynamics of IVR in (nonrotating) OCS have not been investigated. Here, we undertake this task, employing the PES for the ground electronic state reported in Ref. \citenum{Xie} and the multiconfiguration time-dependent Hartree (MCTDH) method\cite{MCTDHPR} for the numerical integration of the TDSE.
%We focus on revealing the quantal mechanism of the process, its role in achieving thermalization, and the energy and time scales involved.
%OCS is an excellent model for studying IVR and thermalization because its classical dynamics are chaotic, which implies thermalization in the classical regime.
%We propagate initial wavepackets in a wide range of excitation energies, including the range in which OCS is, presumably, classically chaotic.

This paper is organized as follows. In Section II we establish the coordinate system and the Hamiltonians employed for the calculations and the analysis, define the choice of initial conditions, introduce the diagnostic tools for analyzing the dynamics, and briefly explain some basic aspects of the computational methodology. In Section III we present and discuss our results for excitation energies ranging from low to near the dissociation threshold. In Section IV we summarize our results, establish the main conclusions of the study, and provide prospects for future research. In an Appendix, we provide a prescription for the calculation of statistical microcanonical averages in terms of a zero-order basis, which eliminates the need to diagonalize the Hamiltonian matrix, in the spirit of Ref. \citenum{Flambaum1997}.

\section{METHODOLOGY}

\subsection{Coordinates and nuclear Hamiltonian}

The analysis of IVR in terms of vibrational energy partitioning among zero-order vibrational DOFs is straightforward if there is a coordinate system in which the coupling among the coordinates is relatively weak. Moreover, if there is a reaction, it is convenient that the reaction coordinate matches as closely as possible a single coordinate. At low energies, normal-mode coordinates are weakly coupled, but at high energies they are strongly coupled due to the anharmonicity of the PES. In this work, we consider excitations ranging from low to high energy, the latter lying near the dissociation threshold. In OCS such threshold occurs along the CS distance, which is contained in both stretching normal modes. Therefore, we adopt a local-mode picture for calculation and analysis of IVR in all energy regimes on a single footing.

Specifically, we employ valence coordinates, which constitute a system of internal coordinates suitable for the implementation of the MCTDH methodology. For OCS, these are the CS distance, $0\leq r_{cs}\leq\infty$, the OC distance, $0\leq r_{oc}\leq\infty$, and the bending angle, $0\leq\theta\leq\pi$. In these coordinates, the kinetic energy operator reads\cite{beck} (we use atomic units throughout)

\begin{widetext}
\begin{eqnarray}\label{Eq:Kop}
\hat{K}=&-&\frac{1}{2m_{cs}}\frac{\partial^{2}}{\partial r^{2}_{cs}}-\frac{1}{2m_{oc}}\frac{\partial^{2}}{\partial r^{2}_{oc}}
-\frac{\cos\theta}{m_{c}}\frac{\partial^{2}}{\partial r_{oc}\partial r_{cs}}+\frac{1}{m_{c}}\left(\frac{1}{r_{cs}}\frac{\partial}{\partial r_{oc}}+\frac{1}{r_{oc}}\frac{\partial}{\partial r_{cs}}\right)\frac{\partial}{\partial\theta}\sin\theta\nonumber\\
&+&\frac{1}{2m_{c}r_{cs}r_{oc}}\left(\cot\theta\frac{\partial}{\partial\theta}\sin\theta\frac{\partial}{\partial\theta}+\frac{1}{\sin\theta}\frac{\partial}{\partial\theta}\sin\theta\frac{\partial}{\partial\theta}\cos\theta\right)\nonumber\\
&-&\left(\frac{1}{2m_{cs}r^{2}_{cs}}+\frac{1}{2m_{oc}r^{2}_{oc}}\right)\frac{1}{\sin\theta}\frac{\partial}{\partial\theta}\sin \theta \frac{\partial}{\partial\theta},
%\end{split}
\end{eqnarray}
\end{widetext}

\noindent where $m_{c}$ is the carbon mass and $m_{cs}$ and $m_{oc}$ are the carbon-sulphur and oxygen-carbon reduced masses, respectively. In this expression, the first and second summands represent the kinetic energies of the CS and OC stretchings, respectively, and the second and third lines together represent the kinetic energy of the bending. The third, fourth and fifth summands constitute kinetic couplings, the third one mainly between the two stretchings, the fourth one mainly between the OC stretching and the bending, and the fifth one mainly between the CS stretching and the bending.
%It is noteworthy that this expression contains terms involving the angular variable not present in the kinetic energy employed in Refs. \citenum{CarterBrumer,*carter83,DavisWagner,davis84,Davis,Martens,Gibson,Chandre,PaskaukasPRL,paskaukas}, although the latter does contain the dominant terms, as will be explained below.

For the potential energy operator we use the electronic ground state PES reported in Ref. \citenum{Xie}, which is represented by means of the Morse-cosine expansion

\begin{eqnarray}\label{Eq:Vop}
V(r_{cs},r_{oc},\theta)&=&\sum_{i,j,k}f_{ijk}y^{i}_{cs}y^{j}_{oc}y^{k}_{\theta},\nonumber\\
y_{l}&=&1-e^{-\alpha_{l}(r_{l}-r_{l,e})},\ \ \ \ l=cs,oc,\nonumber\\
y_{\theta}&=&\cos\theta-\cos\theta_{e},
\end{eqnarray}

\noindent %where $\alpha_{l}$ is the Morse exponential parameter and the subscript $e$ in a coordinate indicates its equilibrium value.
where the geometrical parameters are fixed at the experimentally observed values, $r_{cs,e}=2.9506$ bohr, $r_{oc,e}=2.1849$ bohr, $\theta_e=180^\circ$, the Morse exponential parameters are fixed at $\alpha_{cs}=1.03$ bohr$^{-1}$, $\alpha_{oc}=1.24$ bohr$^{-1}$, and the set of parameters $f_{ijk}$ were optimized employing a self-consistent field-configuration interaction method involving experimentally determined vibrational band origins up to 8000 cm$^{-1}\approx 0.04$ hartree. The optimal $f_{ijk}$ values are tabulated in Ref. \citenum{Xie}.

The accuracy of this PES is guaranteed to be better than $5\times 10^{-5}$ hartree, up to an energy of about 0.04 hartree. However, this PES exhibits a dissociation threshold of 0.21666 hartree, which is about twice the experimental value of 0.099784 hartree. On the other hand, the PES used in Refs. \citenum{CarterBrumer,*carter83,DavisWagner,davis84,Davis,Martens,Gibson,Chandre,PaskaukasPRL,paskaukas} is fitted to reproduce such experimental value, but displays a likely unphysical well at large $r_{cs}$, which supports a large number of closely packed eigenstates.\cite{Gerbasi} In contrast, the PES of Ref. \citenum{Xie} approaches the dissociation threshold monotonically.
Unfortunately, a PES that is highly accurate in the entire energy range from the vibrational ground state up to the dissociation threshold is not available to date.
We consider our choice of PES more convenient, since it affords a realistic model of OCS up to at least 40\% of the dissociation threshold and at high energies it can be expected to produce qualitatively correct results.

%Since the present study spans that range, we consider our choice of PES more appropriate. Therefore, we have a realistic model of OCS up to at least 40\% of the dissociation threshold, while at high energies our results will be of qualitative value only.

\subsection{Zero-order local-mode Hamiltonians and eigenstates}

Both for the setting of initial conditions and the analysis of the dynamics, we define the zero-order Hamiltonian for each local vibrational mode as the sum of its corresponding kinetic energy operator appearing in Eq. (\ref{Eq:Kop}) and the slice of the PES obtained by fixing the coordinates of the other two local modes at their equilibrium positions in Eq. (\ref{Eq:Vop}), namely

\begin{subequations}\label{eq:zeroorderham}
\begin{align}
\hat{H}^{(cs)}_{0}=&-\frac{1}{2m_{cs}}\frac{\partial^{2}}{\partial r^{2}_{cs}}+\sum_{i}f_{i00}y^{i}_{cs},\\
\hat{H}^{(oc)}_{0}=&-\frac{1}{2m_{oc}}\frac{\partial^{2}}{\partial r^{2}_{oc}}+\sum_{j}f_{0j0}y^{j}_{oc},\\
%\begin{split}
\hat{H}_{0}^{(\theta)}=&\frac{1}{2m_{c}r_{cs}r_{oc}}\left(\cot\theta\frac{\partial}{\partial\theta}\sin\theta\frac{\partial}{\partial\theta}+\frac{1}{\sin\theta}\frac{\partial}{\partial\theta}\sin\theta\frac{\partial}{\partial\theta}\cos\theta\right)\nonumber\\
&-\left(\frac{1}{2m_{cs}r^{2}_{cs}}+\frac{1}{2m_{oc}r^{2}_{oc}}\right) \frac{1}{\sin\theta}\frac{\partial}{\partial\theta}\sin\theta\frac{\partial}{\partial\theta}\nonumber\\
&+\sum_{k}f_{00k}y^{k}_{\theta}.
%\end{split}
\end{align}
\end{subequations}

\noindent The zero-order local-mode eigenenergies and eigenstates are given by

\begin{subequations}\label{Eq:estados1D}
\begin{align}
\hat{H}^{(cs)}_{0} | n \rangle &= E^{(cs)}_{n} |n  \rangle,\\
\hat{H}^{(oc)}_{0} | m \rangle &= E^{(oc)}_{m} |m \rangle,\\
\hat{H}^{(\theta)}_{0} | l \rangle &= E^{(\theta)}_{l} | l \rangle,
\end{align}
\end{subequations}

\noindent The eigenstates of the zero-order, or noninteracting, Hamiltonian, $\hat{H}_0=\hat{H}^{(cs)}_{0}+\hat{H}^{(oc)}_{0}+\hat{H}^{(\theta)}_{0}$,
%
%\begin{equation}\label{Eq:nonintHamilt}
%\hat{H}^0=\hat{H}^{0}_{cs}+\hat{H}^{0}_{oc}+\hat{H}^{0}_{\theta},
%\end{equation}
%
are the product states $|n\rangle\otimes|m\rangle\otimes|l\rangle\equiv |n'\ m'\ l'\rangle$.
%
%\begin{equation}\label{eq:prodstates}
%|n\rangle\otimes|m\rangle\otimes|l\rangle\equiv |n'\ m'\ l'\rangle,
%\end{equation}

The full vibrational Hamiltonian can now be written as $\hat{H}=\hat{H}_{0}+\hat{H}_{I}$, where $\hat{H}_{I}$ is the operator that represents the residual interaction, or coupling, between the local modes.

%\begin{equation}\label{Eq:Hamiltdecomp}
%\hat{H}=\hat{H}^{0}_{cs}+\hat{H}^{0}_{oc}+\hat{H}^{0}_{\theta}+\hat{H}_{I}\equiv\hat{H}^{0}+\hat{H}_{I},
%\end{equation}
%\noindent where $\hat{H}^0$ is the zero-order noninteracting Hamiltonian and $\hat{H}_{I}$ is the operator that represents the residual interaction, or coupling, between the local modes.

\subsection{Initial conditions}

We prescribe the initial state as a zero-order product state,

\begin{equation}\label{eq:initial}
|\Psi(0)\rangle=|n'\ m'\ l'\rangle,
\end{equation}

\noindent where the vibrational quantum numbers $n',m',l'$ label the excitations in the CS, OC, and bending modes, respectively. In this work we focus on IVR from initial states that contain an excitation in only one local vibrational mode, i.e., with the quantum number of the initially excited mode larger than 1 and the quantum numbers of the other two modes equal to 0.

Evidently, since the product states (\ref{eq:initial}) are not eigenstates of $\hat{H}$, they are nonstationary states. Therefore, in terms of the eigenstates of the full Hamiltonian, $|\varphi_a\rangle$, their time evolution can be represented as

\begin{equation}\label{eq:eigenexpansion}
|\Psi(t)\rangle=e^{-i\hat{H}t}|n'\ m'\ l'\rangle=\sum_a b_a e^{-iE_at}|\varphi_a\rangle,
\end{equation}

\noindent with $E_a$ being the eigenenergies and $b_a\equiv\langle\varphi_a|n'\ m'\ l'\rangle$. Since the zero-order product states constitute a basis in the vibrational Hilbert space of the molecule, we have the alternative representation

\begin{equation}\label{eq:zeroexpansion}
|\Psi(t)\rangle=\sum_{n,m,l}c_{nml}(t)|n \ m \ l\rangle,
\end{equation}

\noindent with $c_{nml}(t)\equiv\langle n \ m \ l| e^{-i\hat{H}t}|n'\ m'\ l'\rangle$. In both expansions, the summation is understood to encompass both the discrete and continuous parts of the eigenspectrum.

\subsection{Diagnostic tools}

As is usually done, we characterize the energy content of each local vibrational mode by means of the expectation value of the corresponding zero-order Hamiltonian,

\begin{equation}\label{eq:energy}
\langle E^{(i)}\rangle(t)=\langle\Psi(t)\vert\hat{H}^{(i)}_0\vert\Psi(t)\rangle, \ \ i=cs,oc,\theta. 
\end{equation}

\noindent Clearly, the sum of these expectation values cannot equal the expectation value of the full Hamiltonian, $\langle E\rangle=\langle\Psi(t)|\hat{H}|\Psi(t)\rangle$ (which is actually independent of time), since the latter contains the residual interaction. Hence, we also evaluate the expectation value of the operator

\begin{equation}\label{eq:coupling}
\hat{H}_{I}\equiv\hat{H}-(\hat{H}_{0}^{(cs)}+\hat{H}_{0}^{(oc)}+\hat{H}^{(\theta)}_{0})
\end{equation}

\noindent and, as a test of numerical accuracy, check that at all times the sum of these four expectation values equal $\langle E\rangle$.

For elucidating redistribution pathways, we calculate the time-dependent probabilities, or populations, of zero-order product states,

\begin{equation}\label{eq:prodpop}
P_{nml}(t)\equiv|c_{nml}(t)|^2.
\end{equation}

\noindent Note that when $n=n',m=m',l=l'$, Eq. (\ref{eq:prodpop}) is just the survival probability.

Following Wolynes and coworkers,\cite{Wolynes1,Gruebele2} we regard the set of zero-order quantum numbers $n,m,l$ as the ``coordinates" of a three-dimensional lattice, the so-called QNS, with each axis representing the zero-order eigenstates of a corresponding mode and every point $(n\ m\ l)$ representing a zero-order product state.
Hence, the populations (\ref{eq:prodpop}) can be regarded as the occupations of the ``cells" in the QNS.
%This approach allows IVR to be treated in a manner analogous to the theory of Anderson localization of electron motion in disordered semiconductors.\cite{Wolynes1} 
In particular, we wish to determine whether a transition from localized to delocalized energy flow takes place in OCS, as has been revealed for more complex molecules. \cite{Wolynes1,Gruebele2}
Nevertheless, even in this low-dimensional QNS, except at very low energies, a detailed examination of the dynamics of the population flow becomes unwieldy, so one must resort to statistical analyses.\cite{Wolynes1,Gruebele2}
Here, we compute time-averaged populations in the QNS as

\begin{equation}\label{eq:av-prodpop}
\overline{P_{nml}}=\frac{1}{T}\int_{0}^{T}P_{nml}(t) dt,
\end{equation}

\noindent where $T$ is the total propagation time.

Since the dynamics are governed by the contributions of the eigenstates to the nonstationary state [see Eq. (\ref{eq:eigenexpansion})], much insight can be obtained from the so-called energy distribution function, $P(E)=\sum_a |b_a|^2\delta(E-E_a)$.\cite{Dalessio,Kaufman,Santos}
In our case, the direct computation of this function is impractical, so we obtain a good approximation to it via the spectral function, which is defined as the Fourier transform\cite{MCTDHPR,meyer2009}

\begin{equation}\label{eq:spectralfunction}
\sigma(E)=\int_{-\infty}^{\infty}C(t)W(t)e^{iEt}dt, 
\end{equation}

\noindent where $C(t)\equiv\langle n'\ m'\ l'\vert\Psi(t)\rangle$ is the autocorrelation function and $W(t)$ is a window function that takes into account that $C(t)$ is known for the interval $[-T,T]$, with the values for $[-T,0)$ obtained for free by taking advantage of the time-reversal invariance $C(-t)=C^*(t)$. By inserting the expansion (\ref{eq:eigenexpansion}) in Eq. (\ref{eq:spectralfunction}) we obtain

\begin{equation}\label{eq:spectralfunction2}
\sigma(E)=\sum_a |b_a|^2 L(E-E_a), 
\end{equation}

\noindent where $L(E-E_a)$ is a lineshape function with the property $\lim_{T\rightarrow\infty}L(E-E_a)=\delta(E-E_a)$. The explicit forms of $W(t)$ and $L(E-E_a)$ are given in Eqs. (170) and (171) of Ref. \citenum{MCTDHPR}.
%Thus, it is seen that $\sigma(E)$ is the convolution of $P(E)$ with $L(E)$, which is the Fourier transform of $W(t)$.
Therefore, for long enough $T$ the spectral function consists of a series of peaks centered at the eigenenergies and with heights proportional to the spectral weights $|b_a|^2$.

%In classical mechanics, the entropy of a subsystem, which is defined in terms of its reduced probability distribution, constitutes a standard measure of delocalization in phase space. For ergodic (or, more precisely, mixing) systems, this distribution spreads throughout all the energetically available phase space and the entropy reaches a maximum value consistent with the Gibbs distribution at long enough time. For non-ergodic systems, on the contrary, this distribution remains localized in a region of phase space and the entropy does not reach such value.
%In quantum mechanics, the reduced density operator and the von Neumann entropy play analogous roles.\cite{Dalessio} Therefore, we employ the von Neumann entropy as a measure of energy delocalization in QNS, which is defined by

When the entire system is in a pure state, as in our case, the subsystem entanglement entropy plays the role of the thermal entropy.\cite{Huse,Gogolin,Dalessio,Kaufman,Neill,Santos} Hence, this quantity can also serve as a measure of energy delocalization in the QNS. For its definition, we employ the von Neumann entropy

\begin{equation}\label{eq:entropy}
S_{i}(t)=-\tr\rho_{i}(t)\ln\rho_{i}(t)
=-\sum_{\lambda}p^{(i)}_{\lambda}(t)\ln p^{(i)}_{\lambda}(t),
\end{equation}

\noindent where $\rho_{i}(t)$ is the reduced density matrix of the $i$-th ($i=cs,oc,\theta$) zero-order local mode and $p^{(i)}_{\lambda}(t)$ are its eigenvalues.
%When the entire system is in a pure state, as in our case, the von Neumann entropies are also called entanglement entropies.

To help assess to what degree the molecule thermalizes, we compute the populations of zero-order states of each local mode by

\begin{subequations}\label{eq:pobs}
\begin{align}
 p_{n}^{(cs)}(t)&=\langle\Psi(t)|\hat{P}_{n}^{(cs)}
 %| \widetilde{n} \rangle \langle \widetilde{n}
 |\Psi(t)\rangle=\sum_{m,l}P_{nml}(t),\\
 p_{m}^{(oc)}(t)&=\langle\Psi(t)|\hat{P}_{m}^{(oc)}
 %| \widetilde{m} \rangle \langle \widetilde{m}
 |\Psi(t)\rangle=\sum_{n,l}P_{nml}(t),\\
 p_{l}^{(\theta)}(t)&=\langle\Psi(t)|\hat{P}_{l}^{(\theta)}
 %| \widetilde{l} \rangle \langle \widetilde{l}
 |\Psi(t)\rangle=\sum_{n,m}P_{nml}(t),
\end{align}
\end{subequations}

\noindent where $\hat{P}_{n}^{(cs)}=|n\rangle\langle n|\otimes \sum_{m,l}|m\ l\rangle \langle m \ l|$ is the projector over the $n$-th state of the CS mode, acting in the full Hilbert space, and analogously for the other two modes.
[If the subsystem undergoes complete decoherence, $\rho_i$ becomes diagonal with the $p^{(i)}_{\lambda}$ of Eq. (\ref{eq:entropy}) as its elements, which, in turn, are given by Eqs. (\ref{eq:pobs}).]

Furthermore, we calculate statistical microcanonical averages of the zero-order Hamiltonians (\ref{eq:zeroorderham}) by means of

\begin{equation}\label{eq:ensemble}
\langle E^{(i)}\rangle_{mic}=\sum_{n,m,l}\frac{1}{\Omega}\langle n\ m\ l | \hat{H}^{(i)}_{0} | n\ m\ l \rangle,
\end{equation}

\noindent where the sum runs over all $\Omega$ \emph{zero-order} states $| n\ m\ l \rangle$ contained within a narrow energy window centered at $\langle E\rangle$-$\langle \Psi(T)|\hat{H}_{I}|\Psi(T)\rangle$. This procedure avoids the use of the eigenstates of $\hat{H}$, whose calculation is very difficult at high energy. We provide a justification for this prescription in the Appendix.

Given that the initial state is nonstationary, at high excitation energy part of the wavefunction may escape into the continuum along the dissociation coordinate, $r_{cs}$. To avoid the use of a very long computational grid, at the same time properly simulating the continuum, we added to the PES in the asymptotic region of the $r_{cs}$ coordinate an absorbing potential. (Fortunately, the presence of this potential does not spoil the aforementioned time-reversal invariance of the autocorrelation function.\cite{MCTDHPR}) Since such potential causes the norm of the wavepacket to decrease as soon as it reaches the asymptotic region, we evaluate the dissociation probability by means of the prescription

\begin{equation}\label{eq:diss}
P_{D}(t)= 1-\vert\langle \Psi(t)\vert \Psi(t)\rangle\vert^{2}.
\end{equation}

\noindent Due to the decrease in the norm, for the calculation of all kinds of averages we renormalize the wavefunction.

\vspace{0.2cm}

% Eq. (\ref{eq:energy}), populations, Eqs. (\ref{eq:prodpop}) and (\ref{eq:pobs}), and microcanonical averages, Eq. (\ref{eq:ensemble})

\subsection{Computational considerations}

For the numerical integration of the TDSE,

\begin{equation}
\left[i\frac{\partial}{\partial t}-\hat{K} -V(r_{cs},r_{oc},\theta)\right]\Psi(r_{cs},r_{oc},\theta,t)=0,
\end{equation}

\noindent with $\hat K$ and $V$ given by Eqs. (\ref{Eq:Kop}) and (\ref{Eq:Vop}), respectively, we employ the MCTDH algorithm\cite{MCTDHPR,meyer2009} implemented in the Heidelberg package.\cite{paquete} In this algorithm, the wavefunction is represented as a linear combination of Hartree products, each one of them consisting of a product of so-called single-particle functions (SPFs), namely

%\begin{widetext}
\begin{eqnarray}
\Psi(r_{cs},r_{oc},\theta,t)&=&\sum^{n_{cs}}_{j_{cs}=1}\sum^{n_{oc}}_{j_{oc}=1}\sum^{n_{\theta}}_{j_{\theta}=1}A_{j_{cs}j_{oc}j_{\theta}}(t)\varphi^{(cs)}_{j_{cs}}(r_{cs},t)\nonumber\\
&\times&\varphi^{(oc)}_{j_{oc}}(r_{oc},t)\varphi^{(\theta)}_{j_{\theta}}(\theta,t),
\end{eqnarray}
%\end{widetext}

%\begin{eqnarray}
%\Psi(r_{cs},r_{oc},\theta,t)&=&\sum^{n_{1}}_{j_{1}=1}...\sum^{n_{f}}_{j_{f}=1}A_{j_{1}...j_{f}}(t)\prod^{f}_{i=1}\varphi^{(i)}_{j_{i}}(Q_{i},t),
%\nonumber\\
%&\equiv&A_{J}\Phi_{J},
%\end{eqnarray}

\noindent where
%$i=cs,oc,\theta$, $Q_{i}$ is a nuclear coordinate, and 
$n_{i}$ is the number of SPFs for the $i$-th DOF. For each SPF $\varphi^{(i)}_{j_{i}}(q_{i},t)$, in turn, a discrete variable representation (DVR) is used. We chose a harmonic oscillator DVR for the stretching modes and a Legendre DVR for the bending mode.

The Heidelberg MCTDH algorithm requires that also the PES be represented as a linear combination of products of single-variable functions. Expression (\ref{Eq:Vop}) is already in this form.

The absorbing potential mentioned in the previous subsection has the form\cite{MCTDHPR,meyer2009}

\begin{equation}
V_{abs}(r_{cs})=-i\eta(r_{cs}-r_{abs})^2\Theta(r_{cs}-r_{abs}),
\end{equation} 

\noindent where $\Theta(r)$ is the Heaviside step function, $r_{abs}$ is the initial point, and $\eta$ is the strength.
%One must make sure that the parameter $b$ is large enough, so that bound-state contributions to the wavefunction do not get absorbed. Moreover, if the parameter $a$ is not properly tuned the continuum portion of the wavefunction will reflect.
Choosing $r_{abs}=8.0$ bohr, we achieved convergence in the behavior of the dissociation probability (\ref{eq:diss}) with $\eta=0.075$ hartree.

It has been shown that, once the number of SPFs and the size of the DVR bases are optimized, MCTDH is a highly efficient and accurate tool for studying IVR. \cite{MeyerHCF3,Meyertolueno,MeyerHFCO,MeyerDFCO,HFCODFCO}

\section{RESULTS AND DISCUSSION}

\subsection{Zero-order eigenspectra and density of states}

The zero-order potentials, given by the last summands in each of Eqs. (\ref{eq:zeroorderham}), measured from the global minimum of the PES, are displayed in Fig. \ref{enerlocales} up to the dissociation threshold. It is seen that, in this energy range, the anharmonicity of the CS stretching is much stronger than the one of the OC stretching, and the anharmonicity of the bending is barely noticeable. 

Superimposed upon these potentials are shown the corresponding energy spectra [see  Eqs. (\ref{Eq:estados1D})], which were calculated by means of the Lanczos diagonalization method implemented in the Heidelberg package.\cite{paquete} 

%The PES used in Refs. \citenum{CarterBrumer,*carter83,DavisWagner,davis84,Davis,Martens,Gibson,Chandre,PaskaukasPRL,paskaukas}, displays the same qualitative behavior, although its dissociation threshold is 0.099784 hartree, whereas the one of our PES is 0.21666 hartree. Thus, care must be exercised when comparing their results with ours.

\begin{widetext}
\begin{center}
\begin{figure}[ht]
\includegraphics[scale=0.16]{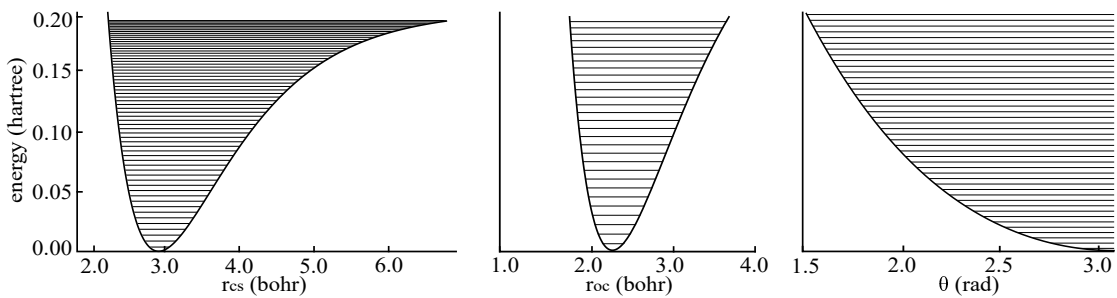}
\caption{Zero-order local-mode potentials and energies. From left to right: CS stretching, OC stretching, and bending.}\label{enerlocales}
\end{figure}
\end{center}
%\end{widetext}

Figure \ref{regime} displays the number of zero-order product states as a function of energy, $n_0(E)$. Due to the residual interaction, any zero-order product state is effectively coupled to a band of the other zero-order product states. Since the zero-order DOS, $dn_0/dE,$ increases with the energy, the density of such band also increases with the energy. This is depicted in Fig. \ref{regime}.

\begin{center}
\begin{figure}[htb]
\includegraphics[scale=0.22]{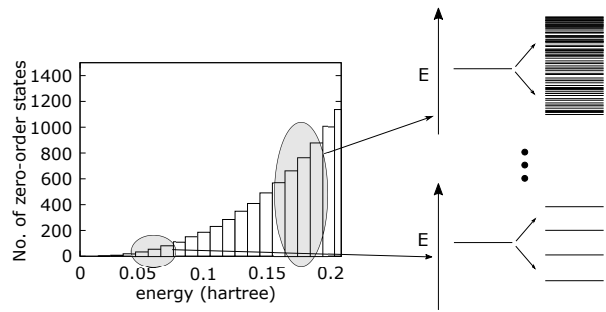}
\caption{Histogram of the number of zero-order states versus energy. The width of the bins is 0.01 hartree. The low- and high-energy regimes are highlighted and the bands to which zero-order states within these regimes are coupled are pictorially illustrated.}
\label{regime}
\end{figure}
\end{center}

\subsection{Energy partitioning}\label{enpart}

We focus on initial excitations localized in so-called edge states, i.e., states with only one zero-order local mode excited at a time [see Eq. (\ref{eq:initial})].
We consider three excitation regimes: low, intermediate, and high, corresponding to energies up to $\sim 15\%$, $\sim 60\%$, and $\sim 90\%$ of the dissociation threshold, respectively.
The expectation values of the energy in each local mode, calculated according to Eq. (\ref{eq:energy}), together with the expectation value of the residual coupling operator given in Eq. (\ref{eq:coupling}), are shown in Fig. \ref{IVR}.
%The top, center, and bottom rows correspond to low-, intermediate-, and high-energy excitations of each local mode, respectively.
It is observed that the magnitude of the expectation value of the residual coupling is comparable to the ones of the local vibrational modes. Thus, a quantitative analysis of IVR in terms of energy partitioning among modes is severely hampered [except at high energy and long time (bottom row), as will be discussed in Subsection \ref{thermalization}]. By examining the behaviors of the contributions to the expectation value of the residual coupling, we determined that, by far, the dominant terms are the kinetic couplings appearing in Eq. (\ref{Eq:Kop}). The reason why these couplings are so strong is that the mass of the carbon nucleus, which appears in the denominator, is comparable to the masses of the oxygen and sulfur nuclei. Moreover, we found that the expectation value of the kinetic coupling between the two stretchings is considerably larger than the one between the bending and the stretchings.

%\small
%\begin{equation}\label{eq:kincoupling}
%-\frac{\cos\theta}{m_{c}}\frac{\partial^{2}}{\partial r_{oc}\partial r_{cs}}+\frac{1}{m_{c}}\left(\frac{1}{r_{cs}}\frac{\partial}{\partial r_{oc}}+\frac{1}{r_{oc}}\frac{\partial}{\partial r_{cs}}\right)\frac{\partial}{\partial\theta}\sin\theta,
%\end{equation}
%\normalsize

%\noindent where the first is the kinetic coupling between the OC and CS stretching modes, and the second is the kinetic coupling between the bending mode and both stretching modes. The reason why this coupling is so strong is that the mass of the carbon nucleus is comparable with the masses of the oxygen and sulfur nuclei. (Since it appears in the denominator, had this mass been larger, the residual coupling would have been smaller.) These terms are indeed included in the kinetic energy used in Refs. \citenum {CarterBrumer,*carter83,DavisWagner,davis84,Davis,Martens,Gibson,Chandre,PaskaukasPRL,paskaukas}.

%\begin{widetext}
\begin{center}
\begin{figure*}[htb]
\includegraphics[scale=0.23]{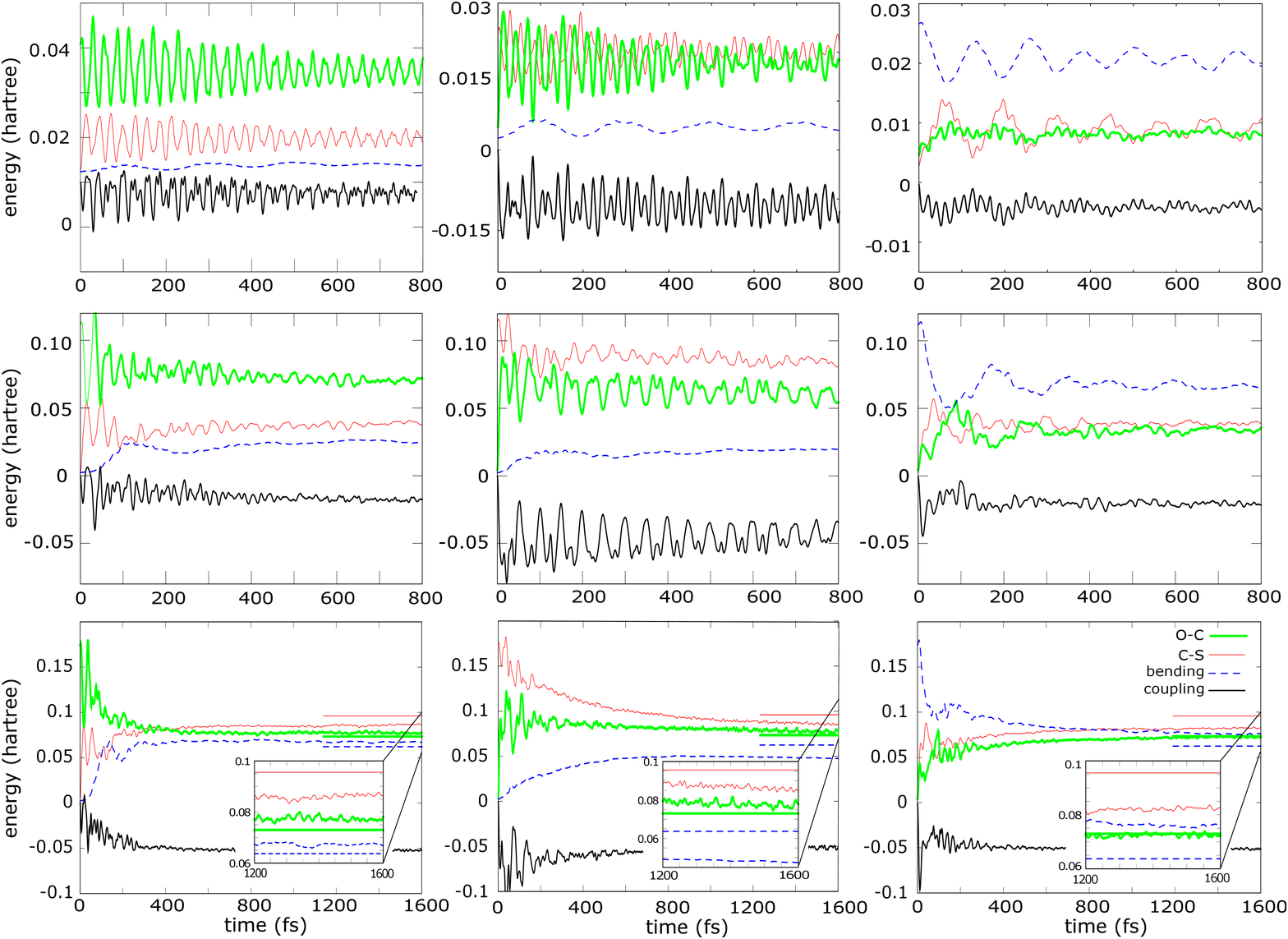}
%{IVRb.eps}
\caption{Evolution of the zero-order local-mode energy expectation values for initial edge states with energies approximately equal to 15$\%$ (top row), 60$\%$ (center row), and 90$\%$ (bottom row) of the dissociation threshold. In each row, from left to right the initial excitation is localized in the OC, CS, and bending mode, respectively. Top row: (0 3 0) (4 0 0) (0 0 5). Center row: (0 13 0) (24 0 0) (0 0 22). Bottom row: (0 22 0) (43 0 0) (0 0 34). The insets in the bottom row compare the energy expectation values with the microcanonical averages.}\label{IVR}
\end{figure*}
\end{center}
%\end{widetext}

Nevertheless, for a qualitative analysis it is still instructive to analyze these plots. At low excitation energy, the dynamics are highly coherent, as revealed by the resilient oscillations in the mean energies of the local modes. In particular, when the bending mode is initially excited, its energy exhibits rather regular oscillations, which mirror the oscillations in the envelope of the CS stretching energy. This can be understood by examining Fig. \ref{enerlocales}, where it can be seen that, at low energy, the DOSs of these two modes are similar, which enables quasi-resonant single-quantum energy exchange between them. Meanwhile, the OC mean energy exhibits much less regular and of much lower amplitude oscillations with about the same frequency as the high-frequency oscillations present in the CS energy. This is the signature of off-resonance energy exchange between the two stretchings, as expected from the fact that the OC DOS is significantly smaller than the one of CS at low energy (see Fig. \ref{enerlocales}). In addition, the energy flow between the two stretchings is much faster than the one between the bending and the CS stretching, which must be due to the fact noted above that the CS-OC coupling is stronger than the stretching-bending one.
The coupling energy exhibits weak envelope oscillations with the same frequency as the oscillations in the bending energy and the low-frequency oscillations in the envelope of the CS stretching energy, which must be related to the slow quasi-resonant energy exchange between these two modes. The coupling energy also exhibits fast oscillations with about the same frequency as the oscillations in the OC stretching energy and the high-frequency oscillations in the CS energy, which must be related to the fast off-resonance energy exchange between these two modes.
The behavior of the mean energies when one of the stretchings is initially excited can be rationalized appealing to the same observations about the DOSs and the intermode coupling strengths. Finally, in all cases at low excitation energy the \emph{net} energy flow remains low. Indeed, at 800 fs most of the energy remains localized in the initially excited mode.

In going from low through intermediate to high excitation energy, the oscillations get increasingly damped with time, a signature of decoherence. In addition, the net IVR becomes more effective, to the point that at high energy and $\sim$800 fs a near equipartition of the excitation energy among the three modes is attained. These behaviors must be related to the rapid increase of the DOS with energy (see Fig. \ref{regime}).

\subsection{Pathways at low energy}

%\begin{widetext}
\begin{center}
\begin{figure}[htb]
\includegraphics[scale=0.35]{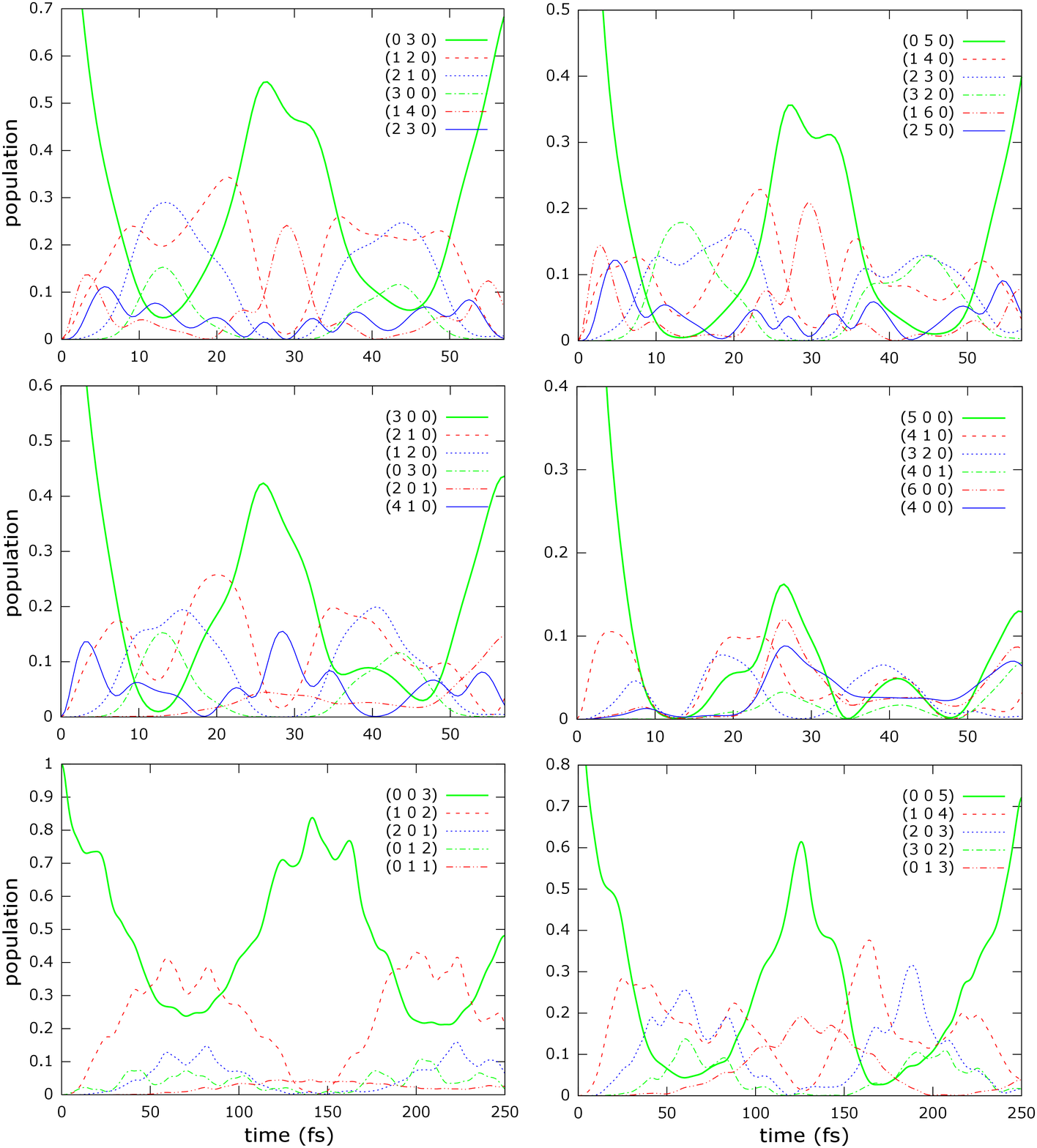}
\caption{Evolution of the zero-order product-state populations for low-energy initial excitations localized in OC mode (top row), CS mode (center row), and bending mode (bottom row). Populations not shown remain one order of magnitude smaller than the ones shown.}\label{paths}
\end{figure}
\end{center}
\end{widetext}

We can obtain a quantitative picture of the IVR mechanism by examining the populations of the zero-order product states, given by Eq. (\ref{eq:prodpop}). In particular, this allows us to infer state-to-state pathways. Figure \ref{paths} shows the evolution of such populations for low-energy initial excitations.

Let us begin by examining the case where the initial edge state contains three quanta in the OC mode. It is observed that at $\sim$3 fs the population has flown through the sequence of transitions (0 3 0) $\rightarrow$ (1 4 0) $\rightarrow$ (1 2 0) $\rightarrow$ (2 3 0) $\rightarrow$ (2 1 0) $\rightarrow$ (3 0 0).
All of these appear to violate conservation of energy, since they are far from resonance. However, it must be kept in mind that the residual coupling between the stretchings is strong, which enables it to act like a source or sink of energy (see Fig. \ref{IVR}).
%Hence, we can envision the single-quantum transitions (0 3 0) $\rightarrow$ (1 2 0), in the first case, and (0 5 0) $\rightarrow$ (1 4 0), in the second case; the two-quantum transitions (0 3 0) $\rightarrow$ (2 1 0), in the first case, and (0 5 0) $\rightarrow$ (3 2 0), in the second case; the three-quantum transitions (0 3 0) $\rightarrow$ (3 0 0), in the first case, and (0 5 0) $\rightarrow$ (2 3 0), in the second case; and the transitions that do not preserve the number of quanta (0 3 0) $\rightarrow$ (1 4 0), in the first case, and (0 5 0) $\rightarrow$ (2 5 0) and (0 5 0) $\rightarrow$ (1 6 0), in the second case.
%The flow cannot be said to be sequential.
%Thus, at low energy the population flow from the OC mode towards the CS mode is, in general, multi-quantum and non-resonant, and towards the bending mode is very small, in accordance with the observations made in the previous subsection.
Within this sequence of transitions we can identify the pathway
%Let us ignore for a moment the states (1 4 0) and (2 3 0) in the first case, and (1 6 0) and (2 5 0) in the second case. Then we would get 
(0 3 0) $\rightarrow$ (1 2 0) $\rightarrow$ (2 1 0) $\rightarrow$ (3 0 0),
%and (0 5 0) $\rightarrow$ (1 4 0) $\rightarrow$ (2 3 0)  $\rightarrow$ (3 2 0), respectively,
which consists of a sequential single-quantum energy transfer between the bonds. An observation of this sort was made by Hutchinson \emph{et al.}\cite{sibert} in a model of the H$_{2}$O molecule with the bending mode frozen. 
To rationalize this behavior, let us analyze the matrix element of the CS-OC kinetic coupling [see Eq. (\ref{Eq:Kop})],

%\begin{widetext}
\begin{eqnarray}\label{Eq:matrixelement1}
&\langle&n \ m \ l|
-\frac{\cos\theta}{m_{c}}\frac{\partial^{2}}{\partial r_{oc}\partial r_{cs}}|n' \ m' \ l'\rangle\nonumber\\
&=&\frac{1}{m_c}\langle n|i\frac{\partial}{\partial r_{cs}} |n' \rangle\langle m|i\frac{\partial}{\partial r_{oc}}|m' \rangle\langle l|\cos\theta|l'\rangle,
\end{eqnarray}
%\end{widetext}

\noindent which is the dominant one. At low energy, the zero-order states of the local stretching modes
%, $|n\rangle$ and $|m\rangle$,
resemble harmonic-oscillator states. Therefore, the first two matrix elements on the right-hand side of the above equation give rise to the approximate selection (propensity) rules $n=n'\pm 1$ and $m=m'\pm 1$. In addition, in the third matrix element $\cos\theta$ varies little within the range of the bending states,
%$|l\rangle$,
producing the propensity rule $l=l'$.
Nevertheless, unlike H$_{2}$O, OCS is a non-symmetric molecule with its two bonds strongly coupled, because the central atom does not have a mass much greater than the masses of the lateral atoms. As a consequence, transitions that do not conserve the number of quanta, like (0 3 0) $\rightarrow$ (1 4 0) and (1 2 0) $\rightarrow$ (2 3 0) become allowed.
Evidently, the bending mode remains unexcited during the first 60 fs, and, in fact, exchanges little energy with the stretchings during the entire dynamics, as seen in Fig. \ref{IVR}, because the kinetic coupling between it and the two stretchings is relatively weaker.
%To account for this behavior, we note that in the third matrix element on the right-hand side of Eq. (\ref{Eq:matrixelement}), the $\cos\theta$ remains roughly constant at low energy, giving rise to the propensity rule $l=l'$.
%population can be transferred between zero-order states that are far from resonance, e.g.  and (0 5 0) $\rightarrow$ (1 6 0). Similar results were found for other initial states.
%At $\sim$13 fs, when the initial state is nearly depleted, the populations follow the order (2 1 0)$>$(1 2 0)$>$(3 0 0)$>$(0 3 0)$>$(1 4 0).
During the first 60 fs, the (0 3 0) population (survival probability) displays Rabi-like oscillations, while the rest of the populations display faster, more complicated, and of lower amplitude oscillations.
%In summary, in this case the population flow involves, by far, only the two stretchings, in accordance with the behavior of the expectation values (see Fig. \ref{IVR}).

In the case where the initial edge state contains five quanta in the OC mode, at $\sim$3 fs the sequence of transitions
(0 5 0) $\rightarrow$ (1 6 0) $\rightarrow$ (1 4 0) $\rightarrow$ (2 5 0) $\rightarrow$ (2 3 0)  $\rightarrow$ (3 2 0) is observed. This observation can be rationalized in a fashion analogous to the one of the previous case. The behavior of the populations up to 60 fs is also analogous to the one of the previous case.

When the initial excitation is localized in the CS mode, the population flow still involves mainly the two stretchings, although in this case there is a sizable flow of population into the bending mode after $\sim$10 fs, which distorts the oscillations in the survival probability from the Rabi-like behavior. This is in accordance with the behavior of the expectation values (see Fig. \ref{IVR}).

Let us now consider the case where the initial edge state contains three quanta in the bending mode.
By examining the behavior of the matrix elements of the kinetic couplings (not shown) and the times at which the states are populated, we inferred that energy flows first towards the CS stretching and later towards the OC stretching, following the  pathways (0 0 3) $\rightarrow$ (1 0 2) $\rightarrow$ (2 0 1) $\rightarrow$ (0 1 1) and (0 0 3) $\rightarrow$ (0 1 2) $\leftarrow$ (1 0 2). Thus, the bending mode first transfers energy directly to the CS mode and then both directly and indirectly (through the CS mode) to the OC mode.
%The CS channel (0 0 3) $\rightarrow$ (1 0 2) $\rightarrow$ (2 0 1) $\rightarrow$ (0 1 1) and the OC channel (0 0 3) $\rightarrow$ (0 1 2), in one case; and the CS channel (0 0 5) $\rightarrow$ (1 0 4) $\rightarrow$ (2 0 3) $\rightarrow$ (3 0 2) and the OC channel (0 0 5) $\rightarrow$ (0 1 3), in the other case.
%Thus, in both cases the energy flow is dominated by single-quantum transitions.
The envelope of the survival probability also shows Rabi-like oscillations, but with a time scale about five times longer than that when the initial excitation is localized in one of the stretching modes, again due to the fact noted above that the bending-stretching kinetic couplings are relatively weaker. These observations are also in accordance with the results displayed in Fig. \ref{IVR}.

In the case where the initial edge state contains five quanta in the bending mode, by an analogous analysis we inferred the pathways (0 0 5) $\rightarrow$ (1 0 4) $\rightarrow$ (2 0 3) $\rightarrow$ (3 0 2) and (0 0 5) $\rightarrow$ (0 1 3).

%Let us consider the matrix element of the bending-CS kinetic coupling:
%\begin{widetext}
%\begin{eqnarray}\label{Eq:matrixelement2}
%&\langle& n \ m \ l |
%\frac{1}{m_{c}r_{oc}}\frac{\partial}{\partial r_{cs}}\frac{\partial}{\partial\theta}{\sin\theta} | n' \ m' \ l' \rangle\nonumber\\
%&=&\frac{1}{m_c}\langle n | i\frac{\partial}{\partial r_{cs}} | n' %\rangle
%\langle m | \frac{1}{r_{oc}} | m' \rangle  \langle l | i\frac{\partial}{\partial\theta}\sin\theta | l' \rangle.
%\end{eqnarray}
%\end{widetext}

As the excitation energy increases, the DOS grows rapidly (see Fig. \ref{regime}), the harmonic approximation breaks down, especially for the CS mode (see Fig. \ref{enerlocales}), and the potential and kinetic couplings of Eq. (\ref{eq:coupling}) become more effective.
Consequently, the existence of simple propensity rules for IVR is precluded and the redistribution pathways become increasingly complicated, as many more transfer channels become open, involving a large number of zero-order states populated through single- and multi-quantum transitions. Here, we do not pursue such a study, since we wish to focus on higher-energy excitations, for which we use a different strategy.

\subsection{Localization $\rightarrow$ delocalization transition}

In this subsection, we focus on initial excitations localized in the CS mode, which suffices for the point we wish to make.
Figure \ref{localization} displays the time-averaged populations in the QNS for excitations approximately at 12\%, 16\%, 26\%, and 36\% of the dissociation threshold, calculated by means of Eq. (\ref{eq:av-prodpop}) with $T=800$ fs.
%(In this context, states with only one local mode excited are called edge states.)
It is observed that at 12\% the populated states are clustered around the initial state, in accordance with our previous conclusion that the IVR pathways at low energy are dominated by sequential single-quantum transitions (see Fig. \ref{paths}) and the net IVR is low (see Fig. \ref{IVR}).
As the excitation energy of the initial state grows, the more diffuse and diluted in the QNS the population gets, suggesting the increasing involvement of fast, nonsequential multi-quantum transitions. This mechanism of flow from the edge towards the interior of the QNS is referred to as vibrational superexchange \cite{leitner}. In previous studies about IVR, it was noted that the population flows mainly within a constant-energy surface in the QNS.\cite{Gruebele2} In OCS, however, because of the strong intermode coupling, the population diffuses in all directions, although not isotropically, since the delocalization is greater into the OC mode than into the bending mode, in agreement with Figs. \ref{IVR} and \ref{paths}.

\begin{widetext}
\begin{center}
\begin{figure}[ht]
\includegraphics[scale=0.4]{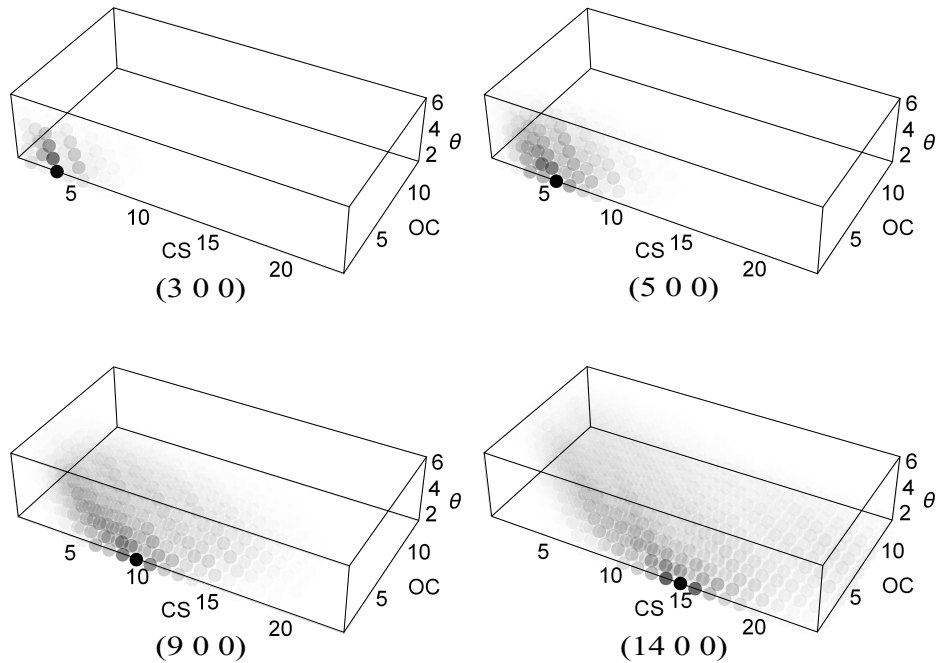}
\caption{Time-averaged populations in the quantum number space for initial excitations localized in the CS mode. The edge states (3 0 0), (5 0 0), (9 0 0), and (14 0 0) correspond to excitations at approximately 12\%, 16\%, 26\%, and 36\% of the dissociation threshold, respectively. The darkness of a circle is proportional to the population of the state, relative to the population of the initial state (black circle).}\label{localization}
\end{figure}
\end{center}
%\end{widetext}

%\begin{figure}
%\begin{center}
%\includegraphics[scale=0.4]{gruebele_picture.eps}
%\caption{Energy shell (``polyad") in the QNS. (From %\cite{Gruebele2}) \textcolor{red}{yo creo que esta parte de esta %imagen definitivamente sobra}}\label{gruebele}
%\end{center}
%\end{figure}
%A standard measure of delocalization in classical mechanics is the entropy, which is defined using the reduced probability distribution in the phase space. In classical ergodic systems, such distribution spreads throughout the available phase space and the entropy reaches the value predicted by the Gibbs distribution. In non-ergodic systems, the probability distribution remains localized in a specific region of the phase space and the entropy has a lower value. In quantum mechanics, an analogous role is played by the reduced density operator\cite{Dalessio}. Hence, we calculate the Von Neumann entropy as a measure of delocalization of the dynamics as
%\begin{equation}
%S_{j} = -\tr \rho_{j}\ln \rho_{j}=-\sum_{i}p^{(j)}_{i}\ln (p^{(j)}_{i}),
%\end{equation}
%where $\rho_{j}$ is the reduced density matrix of the $j-$th vibrational mode (CS, OC or $\theta$) and $p^{(j)}_{i}$ are its eigenvalues. If the system undergoes complete dephasing, $p^{(j)}_{i}$ are the diagonal terms of the reduced density matrix.

We interpret these observations as a continuous transition from local to global energy flow. In the study of Logan and Wolynes,\cite{Wolynes1} the localization $\rightarrow$ delocalization transition occurs when the product of the coupling matrix element and the local DOS reaches a critical value. Here, we have a fixed Hamiltonian, but on average the coupling increases as the energy of the initial state grows (see the second column of Fig. \ref{IVR}), and so does the DOS of the CS local mode (see Fig. \ref{enerlocales}). Not surprisingly, such transition is smooth, since this system has only three DOFs.

Figure \ref{spectra} displays the spectral functions (\ref{eq:spectralfunction2}) corresponding to the initial states of Fig. \ref{localization}, plus two additional initial states of higher energy. It is observed that, concomitantly with the increase of the delocalization in QNS with the excitation energy, there is an increase of the width and the density of the energy distribution. In other words, not too surprisingly, delocalization in QNS is accompanied by delocalization in Hilbert space. Furthermore, it is seen that as the excitation energy increases the envelope of the distribution tends towards a Gaussian centered at the mean energy. These characteristics signal the onset of quantum chaos, as has been observed in other quantum systems.\cite{Santos} Let us recall that the chaotic behavior of IVR in OCS, considering all three internal DOFs, below the dissociation threshold has also been predicted and characterized by means of classical-mechanical methods.\cite{CarterBrumer,DavisWagner,Martens,Chandre,PaskaukasPRL,paskaukas}

%\begin{widetext}
\begin{center}
\begin{figure}[ht]
\includegraphics[scale=0.7]{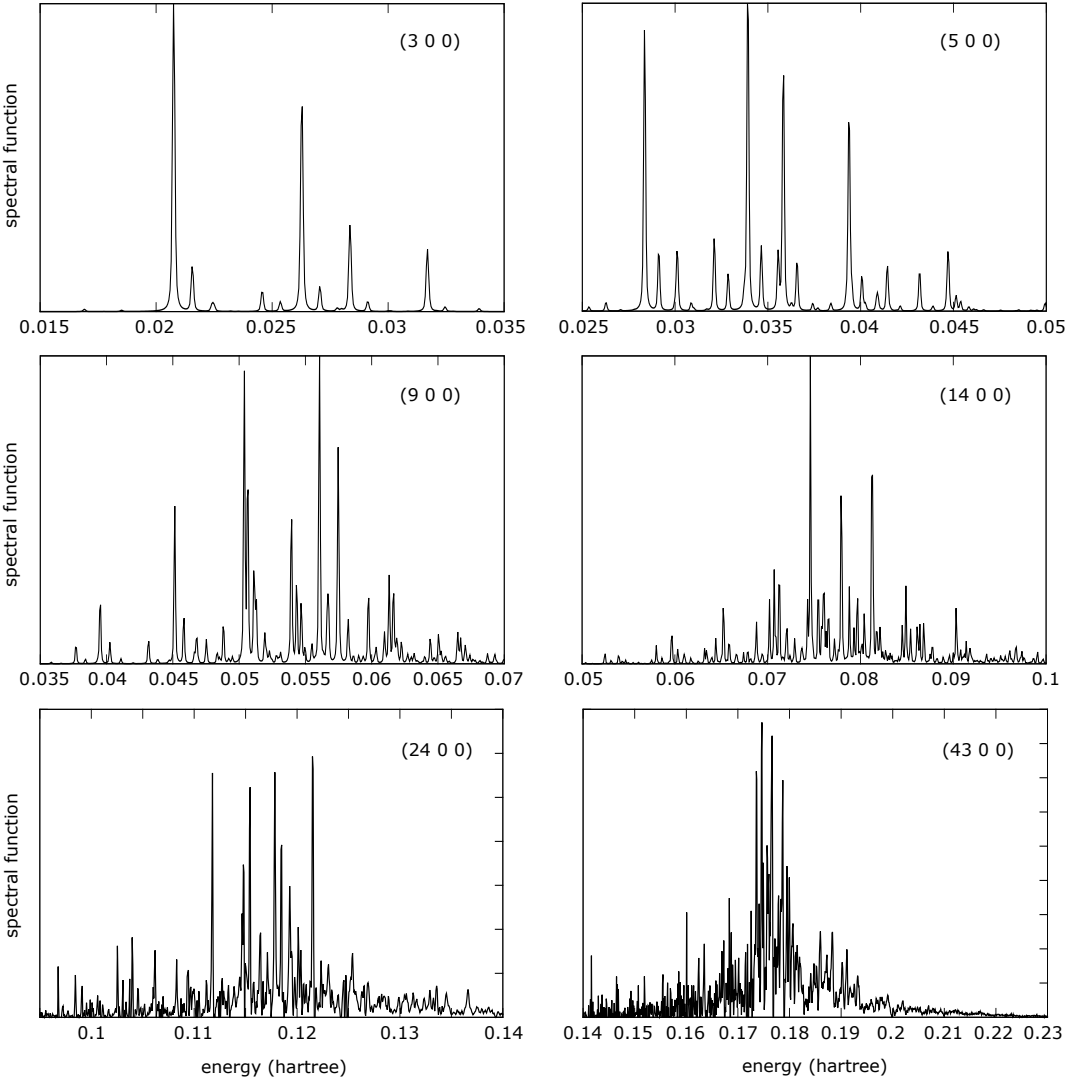}
\caption{Spectral functions (in arbitrary units) for initial excitations localized in the CS mode. The edge states (3 0 0), (5 0 0), (9 0 0), (14 0 0), (24 0 0), and (43 0 0) have mean energies 0.0257, 0.0361, 0.0560, 0.0793, 0.1206, and 0.1799 hartrees, corresponding to approximately 12$\%$, 16$\%$, 26$\%$, 36$\%$, 55$\%$, and 90$\%$ of the dissociation threshold, respectively.}\label{spectra}
\end{figure}
\end{center}
%\end{widetext}

Figure \ref{entropy} shows the time evolution of the entanglement entropy (\ref{eq:entropy}) for each local vibrational mode, for initial excitations at approximately 12$\%$, 16$\%$, 26$\%$, 36$\%$, 55$\%$, and 90$\%$ of the dissociation threshold.
It is observed that this quantity always starts at zero, as expected for an initial zero-order product state, where the modes are not entangled.

Then, at short time ($t\lesssim$10 fs, the order of one vibrational period), in all cases the entropies of the CS and OC modes are nearly equal, while the entropy of the angular mode is much smaller, meaning that the entanglement between the two stretchings is much stronger than between the stretchings and the bending. This is due to the fact that the coupling between the stretchings is stronger than between a stretching and the bending.
In addition, the initial rise of the three entropies is approximately linear, with the rate for the two stretchings being much larger than for the bending. Moreover, in going from low to intermediate excitation energy [(3 0 0)$\rightarrow$(14 0 0)] these rates increase, but at (24 0 0) they have stopped increasing.

%\begin{widetext}
\begin{center}
\begin{figure}
\includegraphics[scale=0.65]{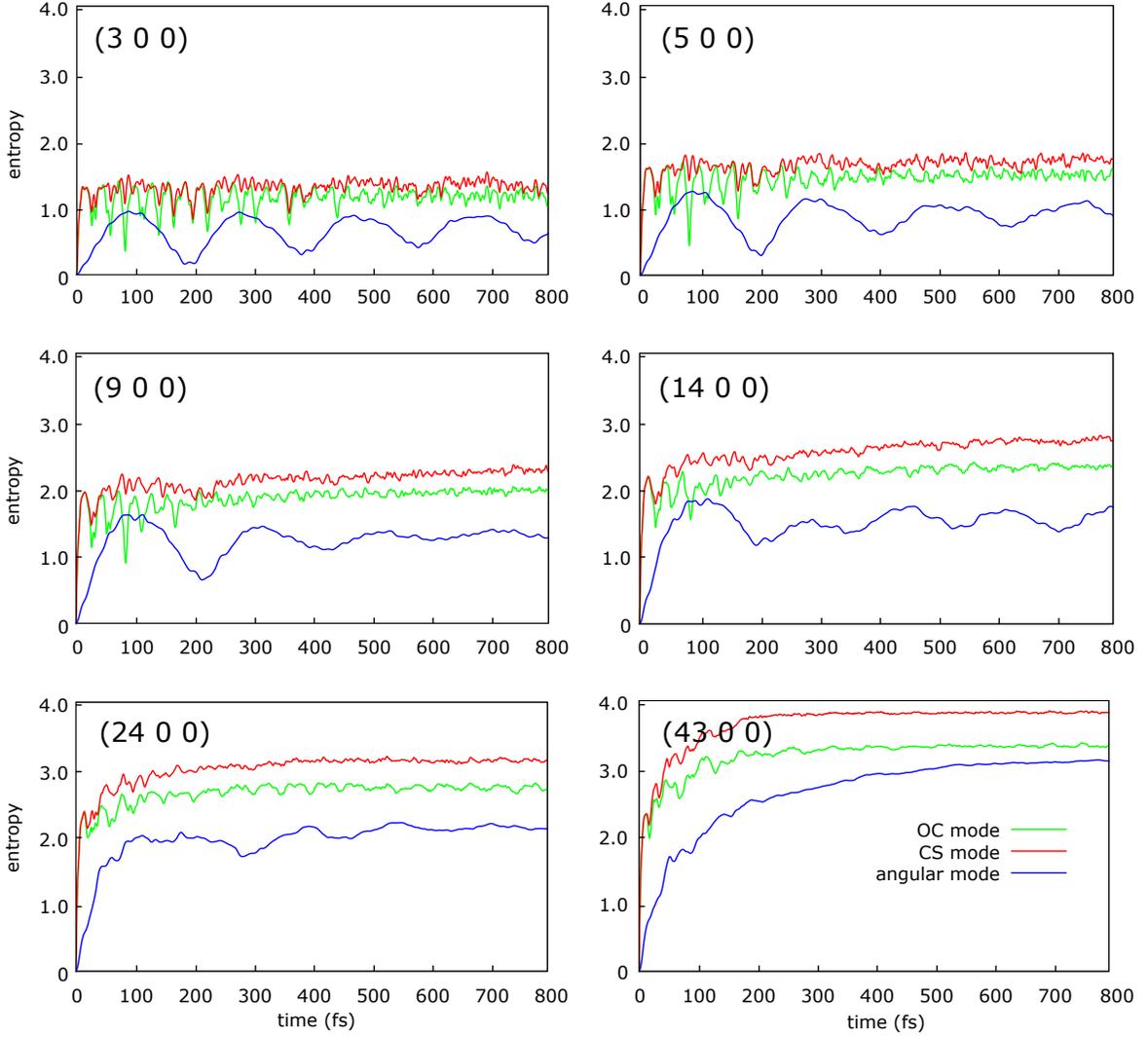}
\caption{Evolution of the entanglement entropies for the same initial states of Fig. \ref{spectra}.}
\label{entropy}
\end{figure}
\end{center}
\end{widetext}

Afterwards, at low energy [(3 0 0) and (5 0 0)] the entropies immediately saturate on average, with fast and irregular oscillations for the stretchings, but with much slower and regular oscillations for the bending. The average value for CS remains slightly above the one for OC and about twice the one for the bending, indicating that the CS-bending entanglement is a little larger than the OC-bending entanglement.
%This is in accordance with the observations in Figs. \ref{IVR} (center frame of the top row), \ref{paths} (second row), and \ref{localization} (top row).
As the energy grows into the intermediate regime [(9 0 0)$\rightarrow$(24 0 0)] the entropies saturate more slowly and the oscillations get increasingly damped with time.
%, in accordance with Figs. \ref{IVR} (center frame of the center row) and \ref{localization} (bottom row).
At high energy [(43 0 0)], at about 250 fs the oscillations are essentially fully damped, and at 800 fs the entropies become steady.
In general, the higher the excitation energy gets the higher the entropy saturation values become at long time.

The short-time linear growth of the entanglement entropy in systems with short-range interactions has been ascribed to the propagation of correlations within an effective causal ``light cone".\cite{Richerme}
In a triatomic molecule, it is obviously the case that the interactions are of short-range, since all the local modes share at least the motion of one nucleus.
%For the CS and OC modes, the fact that such linear growth is so fast must be related to the aforementioned superexchange.

When a system is of finite size, it can be expected that the entropy fluctuate at long times.\cite{Kaufman}
In the present case, the ``sizes" of the subsystems (i.e., the local modes) are fixed, of course. However, what really matters is the size of the effectively available Hilbert space.
Hence, the amplitude of such fluctuations can be related to the sparsity of the energy distribution as a function of excitation energy (see Fig. \ref{spectra}), which, in turn, is influenced by the DOS (see Fig. \ref{regime}): At low excitation energy, where the energy distribution is sparse and the DOS is low, the amplitudes of these oscillations are large, whereas at high excitation energy, where the energy distribution is dense and the DOS is high, the oscillations have practically disappeared.

Since in a pure state the entanglement entropy plays the role of the thermal entropy, which is extensive, it should grow with the subsystem size. It has been proven that for quenched, infinite, and continuous systems such growth should be linear, which is called a volume law.\cite{Eisert}
In the present case, from the long-time vales observed in Fig. \ref{entropy}, we estimate that, roughly, IVR in OCS follows a volume law with respect to the excitation energy. Naturally, the deviations are due to the fact that the available Hilbert space is actually discrete and finite.

From these results, it is seen that the continuous localization $\rightarrow$ delocalization transition exhibited in Fig. \ref{localization} can be characterized by the damping of the temporal entropy oscillations and the increase, on average, of the entropy saturation value as the energy increases. This, in turn, can be associated with the growth in the density and the tendency towards a Gaussian shape centered at the mean energy of the energy distribution.
In OCS, this transition has already begun to occur at relatively low energy [initial state (9 0 0) at 26\% of the dissociation threshold], despite the sparsity of the energy distribution, due to the strong intermode coupling.
By looking back at Figs. \ref{IVR} and \ref{paths}, it can also be verified that such transition is related with the increases in the decoherence and the extent of the net IVR.
%If the system were bigger, the entropy could start decreasing as a result of the energy flow into the other degrees of freedom. However, because this is a very small system, the energy in each mode barely decrease and the entropy reaches a maximum value at long time.

We explore the role of this transition in the ``thermodynamical" behavior of OCS in the next subsection.

%Moreover, in this regime localization breaks down and the DOS is large enough (see Fig. \ref{regime}) to allow the system to reach the statistical limit, as will be seen in the next subsection.

\subsection{Thermalization at high energy}\label{thermalization}

Figure \ref{pop} displays snapshots of the population distributions in each zero-order local mode [see Fig. \ref{enerlocales} and Eqs. (\ref{eq:pobs})], corresponding to excitations at approximately $90\%$ of the dissociation threshold (0.19 hartree) initially localized in one state of the OC stretching (top row), the CS stretching (center row), or the bending (bottom row). (The distributions at $t=0$ are not included, since the one of the initially excited mode consists of a single peak and the ones of the other two modes are identically zero.) The analysis of the evolution of these distributions is aided by an examination of the expectation values of the energy in each mode and of the residual coupling (bottom row in Fig. \ref{IVR} ).

\begin{widetext}
\begin{center}
\begin{figure}[ht]
\includegraphics[scale=0.47]{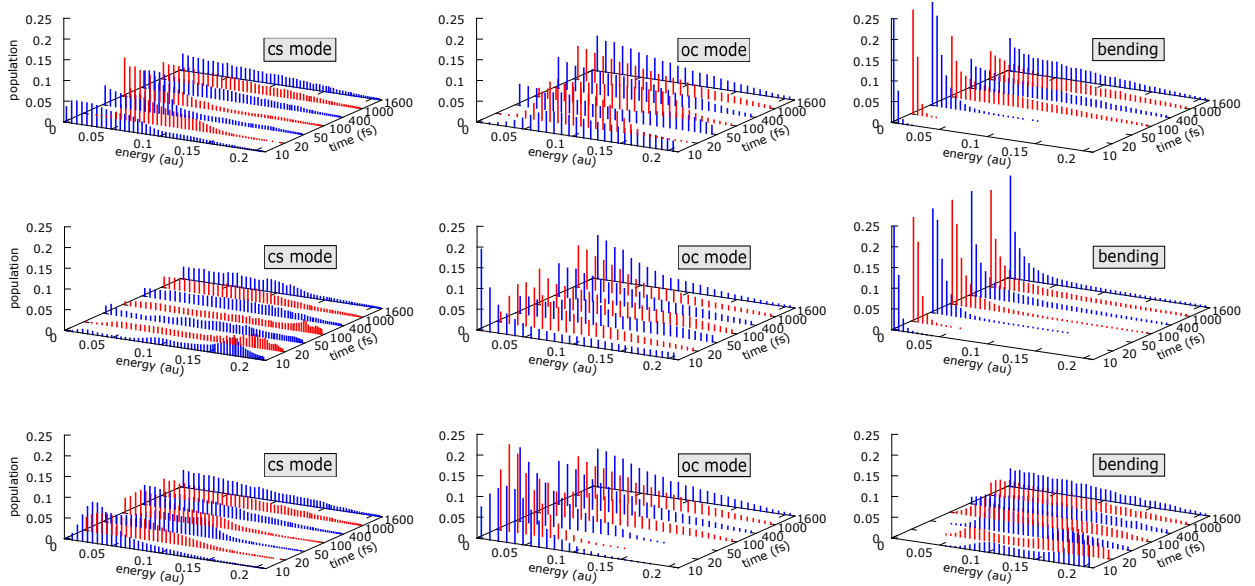}
\caption{Evolution of the population distributions in each zero-order local mode, for initial excitations at approximately $90\%$ of the dissociation threshold. The distributions at $t=0$ are not included (see text). Top row: initial state (0 22 0). Center row: initial state (43 0 0). Bottom row: initial state (0 0 34).}\label{pop}
\end{figure}
\end{center}
\end{widetext}

%\begin{figure}
%\begin{center}
%\includegraphics[scale=0.5]{IVR2b.eps}
%\caption{Evolution of the zero-order local-mode energy expectation values for initial excitations at $90\%$ of the dissociation threshold. Top: initial state (0 22 0). Center: initial state (43 0 0). Bottom: initial state (0 0 34). In the inset, these are compared with the microcanonical averages (straight horizontal lines) in the long-time limit.}\label{IVR2}
%\end{center}
%\end{figure}

After only 10 fs, it is appreciated that the distribution of the initially excited mode becomes very wide and multimodal. In addition, when the initially excited mode is a stretching, the distribution of the other stretching also becomes very wide, while the distribution of the bending becomes very compact. On the other hand, when the initially excited mode is the bending, the distribution of a stretching does not become as wide as when the initially excited mode is the other stretching. Not surprisingly, in all three cases, the distribution of the initially excited mode is slanted toward the high-lying states, while the distributions of the initially unexcited modes are slanted toward the low-lying states. These results are in accordance with the observation pointed out in Subsection \ref{enpart}, that the residual coupling is more effective between the two stretchings than between a stretching and the bending. These distributions can come about only if, at short time and high excitation energy, the energy transfer is dominated by vibrational superexchange, causing the dynamics in the QNS to be diffusive,\cite{Gruebele2} in agreement with the discussion of the previous subsection, in particular with the linear rise of the entanglement entropies.
%This mechanism can be visualized as a very fast flow from the edge towards the interior of the quantum vibrational state space, which is sometimes referred to as vibrational superexchange \cite{leitner}.

As time passes, the distribution of an initially excited stretching sloshes back and forth until finally settling down, whereas the distribution of the initially excited bending spreads without sloshing. On the other hand, the distributions of the initially unexcited modes remain largely slanted towards the low-energy region, especially the one of the bending mode. By 400 fs most distributions have more or less already settled down to decaying shapes. Therefore, for the bending mode the approach to the final distribution is adiabatic-like, whereas for the stretching modes it is clearly not.
All this is in accordance with the behavior of the expectation values, where the ones of the initially excited OC and CS stretchings decay with fast coherent oscillations until about 250 fs, whereas the one of the initially excited bending decays more smoothly. Also, the expectation values of the initially unexcited stretchings display fast oscillations until about 250 fs, while the one of the initially unexcited bending rises much more smoothly.
Beyond $\sim$250 fs, as decoherence takes over, the oscillations get strongly damped and the expectation values begin to settle down, until at 800 fs they have become very steady, with very feeble fluctuations.
The expectation value of the residual interaction becomes steady as well at $\sim$400 fs in the three cases.
The entanglement entropies are seen to behave in concord (last frame in Fig. \ref{entropy}), as the entropies of the CS and OC modes exhibit fast and irregular oscillations up to $\sim$250 fs, when they attain essentially constant values, while the entropy of the bending rises more slowly and smoothly.
%These observations suggest that at long time and high excitation energy the stretching modes still exchange energy through multi-quantum transitions, whereas the bending mode does it largely through sequential single-quantum transitions.

%When the excitation is initially localized either in the OC or CS mode, the angular mode population distribution acquires an exponential-like shape that widens with time. This indicates the transfer mechanism may not be through multi-quantum transition but rather sequential single-quantum transitions from the edge to the interior of the quantum vibrational state space. This could be attributed to the weaker coupling between the angular mode and the stretching modes. although stretching modes shows recurrences at 50 fs, the dynamic becomes diffusive at long time and populations draws exponential a decaying function. This final distributions do not exactly match an exponential form, but this does not imply that the system does not reach a thermal state, as the interaction between each vibrational mode is strong. 
%Indeed, there is a strong dephasing despite the small size of the environment, as a result of the large number of eigenstates involved in the dynamics.

If each local mode is regarded as an open system, with the environment comprising the other two local modes, it is natural to ask if the system thermalizes after a long time. It is well known that a quantum system in contact with a reservoir eventually relaxes into a steady state, where expectation values of system operators become time-independent.\cite{fleming} But a steady state is not necessarily a thermalized state; the latter requires that these expectation values attain the mean values predicted by statistical ensembles.\cite{Huse,Gogolin,Dalessio}
%In Fig. \ref{IVR2}, it is observed that at 1600 fs the expectation values of the energy in each local mode indeed have become steady, with very feeble fluctuations. Correspondingly, at this time the population distributions of Fig. \ref{pop} have settled to constant forms. 
To determine whether the molecule has become vibrationally thermalized, we estimated the microcanonical averages of $\hat{H}_{0}^{(cs)}$, $\hat{H}_{0}^{(oc)}$, and $\hat{H}_{0}^{(\theta)}$ by means of Eq. (\ref{eq:ensemble}). The insets in the bottom row of Fig. \ref{IVR} compare these averages with the corresponding expectation values at long time.
%During the first 200 fs, dynamics is characterized by a highly coherent energy transfer between each local mode. At long time, because the environment as well as the propagation time is finite, we still have time-dependent off diagonal terms in the density matrix that cause oscillations in the expected values and populations. However, contrary to results shown in Figure \ref{IVR}, this coherences are negligible. Indeed, there is a strong dephasing despite the small size of the environment, as a result of the large number of eigenstates involved in the dynamics. 
It can be appreciated that the long-time expectation values, in general, deviate noticeably, but not greatly, from the microcanonical averages. Overall, the smaller deviations are displayed by the OC mode.

An additional criterion for thermalization is the subsystem independence of initial conditions\cite{Gogolin}, i.e., the absence of mode specificity.\cite{Baer-Hase} It is seen that in the three cases of the bottom row of Fig. \ref{IVR} (which have very close, not identical, excitation energies), the long-time mean energy of OC fulfills this condition very closely, the one of CS deviates a little more, and the one of the bending shows the larger deviations. Even the coupling thermalizes closely, in this sense, as has been analogously observed in other systems.\cite{Kaufman}
%it is evident that there is a general deviation, and it is unlikely for these values to be reached in a longer time, as they are already steady.
Hence, we conclude that, at least for the cases considered, OCS nearly thermalizes. In particular, the local modes that most and least closely thermalize are the OC stretching and the bending, respectively. Even though OCS possesses only three internal degrees of freedom, this quasi-thermalization is possible thanks to the large effective size of the available Hilbert space at high energy.
%(It is well to remember that quantum-mechanical evolution is periodic, so the wavefunction will eventually, after a very long time, evolve into a state identical to the initial one, except for numerical errors of the propagation method.)

\begin{figure}[ht!]
\includegraphics[scale=0.67]{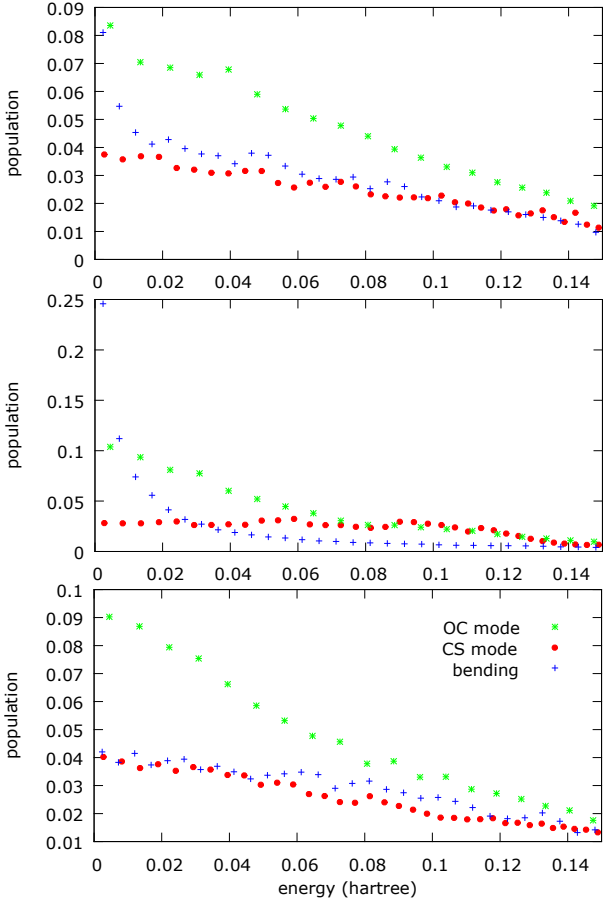}
\caption{Long-time ($t=1.6$ ps) zero-order population distributions of the CS, OC, and bending modes. Top: initial state (0 22 0). Center: initial state (43 0 0). Bottom: initial state (0 0 34).}\label{boltzmann}
\end{figure}

Figure \ref{boltzmann} compares the local-mode population distributions at long time.
Even if the system were fully thermalized, these distributions would not need to attain Boltzmann forms, due the presence of the strong intermode coupling.\cite{fleming} %However, one might expect that when such coupling is relatively ineffective on a particular mode, the corresponding population distribution might be closely Boltzmann-like. Thus, it might be tempting to guess that this is what happens to the bending when the initial excitation is localized in one of the stretchings (top and center rows in Fig. \ref{pop}), since its energy expectation value rises in an adiabatic-like fashion up to its long-time value (bottom row in Fig. \ref{IVR}). Nevertheless, we found that in these cases the final distributions of this mode fit a Boltzmann distribution very poorly.
We found that the distribution that most closely fits a Boltzmann distribution is the one of the OC stretching when the initially excited mode is the CS stretching (center frame in Fig. \ref{boltzmann}). Interestingly, this is the mode that most closely thermalizes according to the criteria of the previous two paragraphs. The Boltzmann fit of this distribution produced a temperature of 18000 K (which corresponds to $k_BT\sim 0.06$ hartree). It is not surprising that the molecule becomes so vibrationally hot, since the initial excitation is so high.

%We compare the populations in each local mode at the last propagation time with the Boltzmann distribution in Figure \ref{boltzmann}.

\subsection{Dissociation}

In the last frame of Fig. \ref{spectra}, it can be appreciated that at $\sim$90\% of the dissociation threshold the tail of the spectral function penetrates slightly into the continuum (recall that the dissociation threshold is at 0.22 hartree). Hence, there is a small probability that the CS bond dissociates. We wish to find out if this dissociation is mode specific and whether it occurs before or after the vibrational quasi-thermalization is attained.

\begin{figure}[ht!]
\begin{center}
\includegraphics[scale=0.6]{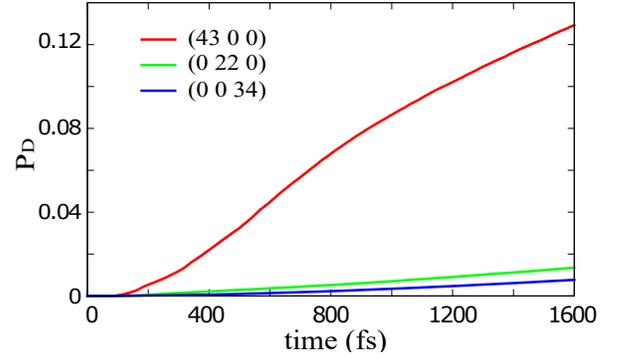}
\caption{Evolution of the CS bond dissociation probability for initial excitations at $\sim$90\% of the dissociation threshold.}\label{dissociation}
\end{center}
\end{figure}

Figure \ref{dissociation} displays the dissociation probability as a function of time, evaluated by means of Eq. (\ref{eq:diss}), for the three cases of Figs. \ref{pop} and \ref{boltzmann} and the bottom row of Fig. \ref{IVR}. Clearly, the time at which the CS bond begins to break is earlier when the initial excitation is localized in that mode ($\sim$100 fs) than in any of the other two modes ($\sim$200 fs). At this time, the dynamics are still largely coherent and the IVR is still far from over. Moreover, the rate of dissociation when the initially excited mode is the CS is much higher than the one when the initially excited mode is the OC or the bending. (Naturally, the dissociation rate is small because the mean energy of the initial wavepacket lies below the threshold.) In the first case, the behavior becomes Golden-Rule-like (approximately linear) at a long time ($\sim$1000 fs), whereas in the other two cases it is Golden-Rule-like essentially since the beginning. 
Hence, it is apparent that, at least for the initial states considered, OCS exhibits a strongly non-RRKM behavior.
In the classical studies reported in Refs. \citenum{Martens,PaskaukasPRL,paskaukas}, the slow IVR in OCS is attributed to bottlenecks in phase space.

\section{SUMMARY, CONCLUSIONS AND OUTLOOK}

%We have performed a rather extensive, although far from exhaustive, fully quantum-mechanical study of the dynamics of IVR in OCS, in terms of zero-order local modes. We have considered only initial excitations localized in one mode.

The first thing we have learnt is that the mean intermode coupling is mainly kinetic and strong, which severely hampers a quantitative analysis in terms of the expectation values of the zero-order Hamiltonians. In addition, such coupling is more effective between the two stretchings than between a stretching and the bending.
%Hence, for any molecule it is imperative to examine the behavior of the expectation value of the residual interaction operator before drawing any conclusions about energy partitioning.
However, these quantities are still very helpful when used in conjunction with more quantitative diagnostic tools, like time-dependent populations of zero-order product states, time averages thereof and their visualization in QNS, energy distributions, time-dependent entanglement entropies, time-dependent population distributions in zero-order local modes, microcanonical averages, and dissociation probabilities.

We have identified three regimes according to the level of excitation: low-energy, intermediate-energy, and high-energy, corresponding, very roughly, to up to 15\%, 60\% and 90\%, respectively, of the dissociation threshold, which occurs along the CS coordinate.
In the low-energy regime, where the PES is nearly harmonic, the local dynamics are highly coherent, involving mostly single-quantum quasi-resonant and off-resonance energy transfer between modes, which makes the identification of IVR pathways and propensity rules straightforward. Nevertheless, the net energy flow is low, which causes the time-averaged zero-order populations to remain localized around the initial state in the QNS.
Concomitantly, the energy distributions are sparse and the entanglement entropies quickly saturate on average.

In the intermediate-energy regime, where the DOS is larger, decoherence becomes manifest, which can be associated with so-called vibrational superexchange, i.e., fast multi-quantum transitions. Consequently, many transfer channels come into play, making the identification of IVR pathways very cumbersome.
Since now the IVR process is more effective, the time-averaged zero-order population becomes more diffuse and diluted in the QNS.
Besides, the energy flow occurs within and off the energy shell and anisotropically, due to the abovementioned characteristics of the coupling.
The energy distributions become wider and denser, the oscillations in the entanglement entropies begin to get damped, and the saturation of these entropies occurs more slowly and to higher values.
These observations signal the appearance of a continuous localization $\rightarrow$ delocalization transition in the energy flow.

In the high-energy regime, at short time the energy transfer is dominated by vibrational superexchange, causing the dynamics in the QNS to be diffusive, albeit still largely coherent. However, after $\sim 250$ fs local decoherence finally takes over.
%Now, the stretching modes still exchange energy through multi-quantum transitions, whereas the bending mode does it mainly through sequential single-quantum transitions.
Now, the envelope of the energy distribution assumes a Gaussian-like shape centered at the mean energy, which is the signature of quantum chaos.
Correspondingly, at $t>800$ fs the entanglement entropies become very steady with even higher values.
In addition, we determined that, roughly, the long-time entanglement entropies follow a so-called volume law with respect to the excitation energy.
%Hence, the continuous localization $\rightarrow$ delocalization transition can be characterized by the damping of the entropy oscillations and the increase, on average, of the entropy values.
Furthermore, at $t>800$ fs a quasi-equilibrium state is reached, in the sense that the mean energies of the local modes attain steady values close to the microcanonical averages, with little mode specificity.
%it is surprising that a system with only three degrees of freedom can approach thermodynamic behavior so closely (quasi-thermalization).
Due to the strong intermode coupling, the population distributions do not reach Boltzmann forms, in general.
%, except for the OC mode when the initially excited mode is the CS, in which case a relatively good fit to a Boltzmann distribution with a quasi-temperature of 18000 K is achieved.
%Moreover, for the bending mode the approach to the final distribution is adiabatic-like, whereas for the stretching modes it is not.
Finally, we found that the dissociation of the CS bond begins well before IVR is over, and that it is highly mode-specific, which means that OCS exhibits a strongly non-RRKM behavior, in agreement with observations made by other authors employing classical-mechanical concepts and methods.

We hope to have shed light on the longstanding problem of the quantum vibrational dynamics of the OCS molecule, as a prototype of a few-DOF system with strong internal couplings. In particular, we hope to have convincingly shown that such system at intermediate and high energies can display phenomena, like a localization $\rightarrow$ delocalization transition and a quasi-thermalization, usually ascribed to more complex systems.

This work may be extended in several directions, for example: First, we recognize that some of the initial states chosen may be difficult, or even impossible, to access experimentally, especially at high energy. Thus, a hard challenge for the near future is to apply optical control schemes\cite{Brumershapiro2} to the preparation of specific states in any energy range. Second, the consideration of initial zero-order states with energies above the dissociation threshold, which is necessary for a deeper understanding of the relationship between the IVR and unimolecular decay processes.\cite{Baer-Hase} Third, devising initial conditions pertaining to thermal activation,\cite{Baer-Hase} i.e., that simulate states created by a collision of the molecule with another molecule of the environment. Fourth, a more complete study of the localization $\rightarrow$ delocalization transition and the associated onset of quantum chaos, considering more general types of excitations. Finally, we would like to investigate if this molecule can undergo a dynamical quantum localization $\rightarrow$ delocalization phase transition, as has been observed in quantum simulators.\cite{Heyl}

\section*{ACKNOWLEDGEMENTS}

\noindent This work was supported in part by Colciencias through project 211765842856. We are grateful to G. A. Zapata for instructive discussions and C. A. Arango for a critical reading of the manuscript.

\section*{APPENDIX: CALCULATION OF MICROCANONICAL AVERAGES USING THE ZERO-ORDER BASIS}

The microcanonical ensemble is an equal-probability mixture of all the eigenstates, $|\varphi_{a}\rangle$, of the Hamiltonian of the system, $\hat{H}$, whose energies, $E_{a}$, lie within an interval around the mean energy,
$\langle E\rangle=\langle\Psi(t)|\hat{H}|\Psi(t)\rangle$. The probabilities are given by

\begin{equation}\label{eq:microprob}
P_{a}=\begin{cases}
1/\Omega & \text{ if } |E_{a}-\langle E \rangle| \leq \delta E\\ 
0 & \text{ if } |E_{a}-\langle E \rangle|> \delta E
\end{cases}
\end{equation}

\noindent where $\Omega$ is the number of eigenstates in an arbitrary interval $\delta E$ around $\langle E\rangle$ (see the Supplementary Materials in Ref. \citenum{Kaufman}). Thus, strictly speaking, to calculate the microcanonical average of an observable $\hat{O}$,

\begin{equation}\label{eq:micromean}
\langle\hat{O}\rangle_{mic}=\sum_{a}P_{a}\langle\varphi_{a} |\hat{O}|\varphi_{a}\rangle=\sum_{a\in\delta E}\frac{1}{\Omega}
\langle\varphi_{a}|\hat{O}|\varphi_{a}\rangle,
\end{equation}

\noindent the eigenspectrum of $\hat{H}$ within $\delta E$ must be known, which technically can be very hard or even impossible. Therefore, it is convenient to expand the eigenstates in a zero-order basis, $|\varphi_a\rangle=\sum_{i}c^{a}_{i}|\psi_{i}\rangle$, so that

\begin{equation}\label{eq:micromeanbasis}
\langle\hat{O}\rangle_{mic}=\sum_{a\in\delta E}\frac{1}{\Omega}\sum_{i,j}c^{a}_{i}c^{*a}_{j}\langle \psi_{j} |\hat{O}|\psi_{i}\rangle,
\end{equation}

\noindent
%where we have used $|\varphi_a\rangle=\sum_{i}c^{a}_{i}|\psi_{i}\rangle$.
In the special case that $\hat{O} |\psi_{i}\rangle=o_{i}|\psi_{i}\rangle$, we get

\begin{equation}\label{eq:micromeanspecial}
\langle \hat{O} \rangle_{mic}=\sum_{a\in\delta E}\left(\sum_{i}\frac{1}{\Omega} |c^{a}_{i}|^{2}\right)o_{i}.
\end{equation}

\noindent To use this expression, we still must know the distribution of the weights $|c^{a}_{i}|^{2}$
%in the representation of $|\varphi_{a}\rangle$ in the zero-order basis,
and the observable of which the zero-order basis is an eigenbasis must commute with $\hat{O}$.
%(e. g. $\hat{H}^{0}_{cs}$, $\hat{H}^{0}_{oc}$ and $\hat{H}^{0}_{\theta}$).
For us, it is convenient to choose as zero-order basis the eigenstates of the noninteracting Hamiltonian $\hat{H}_{0}=\hat{H}_{0}^{(cs)}+\hat{H}_{0}^{(oc)}+\hat{H}_{0}^{(\theta)}$.

According to a study of the structure of chaotic eigenstates,  \cite{Flambaum1997} the weight distribution of a single eigenstate $|\varphi_{a}\rangle$ in the $|\psi_{i}\rangle$ basis is a Gaussian-like function. Therefore, the weight distribution of a range of eigenstates $\{|\varphi_{a}\rangle\}$ in the $|\psi_{i}\rangle$ basis is a sum of overlapping Gaussian-like functions. Since for the calculation of the statistical average (\ref{eq:micromeanspecial}) the precise shape and width of the latter distribution are immaterial, we assume it to be a narrow rectangular window. We illustrate this pictorially in Fig. \ref{App_A}.

\begin{widetext}
\begin{center}
\begin{figure}[ht]
\begin{center}
\includegraphics[scale=0.8]{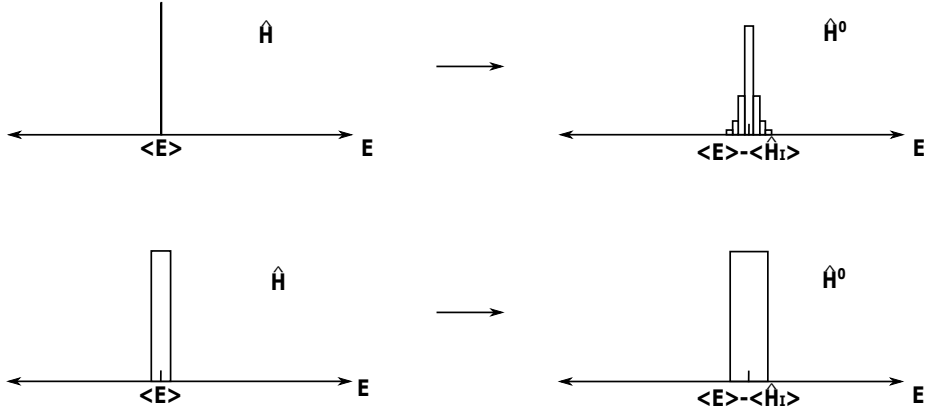}
\caption{Weight distributions in the $\hat{H}$ and $\hat{H}_{0}$ eigenbases. Top: Eigenstate with energy $E_a=\langle E \rangle$. Bottom: Microcanonical ensemble of eigenstates with energies around $\langle E \rangle$.}\label{App_A}
\end{center}
\end{figure}
\end{center}
%\end{widetext}

Nevertheless, the centers of the weight distributions in the eigenstate and  zero-order bases are not the same. To see this, we evaluate the microcanonical average of the energy as follows:

%\begin{widetext}
\begin{eqnarray}\label{eq:meanEmicro}
\langle E\rangle_{mic}&=&\sum_{a}P_{a}\langle \varphi_{a}|\hat{H}|\varphi_{a}\rangle=\sum_{a\in\delta E}\frac{1}{\Omega} \langle \varphi_{a} |\hat{H}_{0}+\hat{H}_{I}|\varphi_{a}\rangle\nonumber\\
%&=\sum_{a\in \Omega}\frac{1}{\Omega} \left(\langle \varphi_{a} |\hat{H}^{0}|\varphi_{a}\rangle +\langle \varphi_{a} |\hat{H}_{I}|\varphi_{a}\rangle\right) \nonumber \\
&=&\sum_{a\in\delta E}\frac{1}{\Omega}\left(\sum_{i,j}c^{a}_{i}c^{*a}_{j}\langle \psi_{j} |\hat{H}_{0} |\psi_{i}\rangle + \langle \varphi_{a}| \hat{H}_{I} |\varphi_{a}\rangle\right)\nonumber\\
&=&\sum_{a\in\delta E}\left(\sum_{i}\frac{1}{\Omega} |c^{a}_{i}|^{2}\right)E_{i}+\sum_{a\in\delta E}\frac{1}{\Omega} \langle \varphi_{a}| \hat{H}_{I} |\varphi_{a}\rangle,
\end{eqnarray}
\end{widetext}

\noindent where $E_i$ are the eigenenergies of $\hat{H}_0$ and we have used Eqs. (\ref{eq:micromean})-(\ref{eq:micromeanspecial}).
%Let us suppose for a moment that the zero-order states were the exact eigenstates of $\hat{H}$. Then,
%On the other hand, the microcanonical average of the zero-order energy in the zero-order basis is given by
%
%\begin{equation}\label{eq:micromeanapprox}
%\langle E^0\rangle_{mic}=\sum_{i}P_{i}\langle\psi_{i} |\hat{H}^{0}|\psi_{i}\rangle=\sum_{i\in\delta E^0}\frac{1}{\Omega'}E^{0}_{i},
%\end{equation}
%
%\noindent where $\Omega'$ is the number of zero-order states in an arbitrary interval $\delta E^{0}$ around $\langle E^{0}\rangle$. Hence, the first summand in the last line of Eq. (\ref{eq:meanEmicro}) can be interpreted as $\langle E^0\rangle_{mic}$ by identifying $\Omega'\equiv\left(\sum_{a\in\delta E}\frac{1}{\Omega}|c^{a}_{i}|^{2}\right)^{-1}$.
%
%\begin{equation}
%\langle E \rangle=\sum_{a\in \Omega}\frac{1}{\Omega}E_{a}=\sum_{i}^{\Omega'}\frac{1}{\Omega'}E_{i} + \langle\hat{H}_{I}\rangle_{micro}.
%\end{equation}
%
In the last line of this equation, the first summand is the microcanonical average of $\hat{H}_0$ in the zero-order basis [see Eq. (\ref{eq:micromeanspecial})] and the second summand is the microcanonical average of $\hat{H}_I$ in the eigenstate basis [see Eq. (\ref{eq:micromean})].
By moving $\langle\hat{H}_I\rangle_{mic}$ to the left-hand side, we see that the energy window for the calculation of $\langle E_0\rangle_{mic}$ is positioned at $\langle E\rangle_{mic}-\langle\hat{H}_I\rangle_{mic}$.
In other words, this equation indicates that we can obtain $\langle E\rangle_{mic}$ from $\langle E_0\rangle_{mic}$, but with the energy window of zero-order eigenstates shifted by the quantity $\langle\hat{H}_I\rangle_{mic}$.
%in the eigenstate basis, can be interpreted as a shift of the energy window used to calculate $\langle E^0\rangle_{mic}$
%in the zero-order basis
%with respect to the energy window in the eigenstate basis. 
(In thermodynamic systems, where the system-bath coupling is commonly small, this shift is negligible.)

The key of our approach to avoid the explicit computation of the latter average is the assumption that the expectation value of the residual coupling thermalizes, which is plausible according to the discussion in Subsection \ref{thermalization} (see the bottom row of Fig. \ref{IVR}). Therefore, we make the identification

\begin{equation}
\langle\hat{H}_{I}\rangle_{mic}=\sum_{a\in\delta E}\frac{1}{\Omega} \langle \varphi_{a}| \hat{H}_{I} |\varphi_{a}\rangle = \langle \Psi(T)|\hat{H}_{I}|\Psi(T)\rangle,
\end{equation}

\noindent where $T$ is a long enough time.
%We must emphasize that this estimation does not require the assumption that the system be thermalized at long time, but only that the expectation value attains an approximately constant value, as it is indeed observed in the bottom row of Fig. \ref{IVR}.
Consequently, for the calculation of microcanonical averages we employ Eq. (\ref{eq:micromeanspecial}), with an energy window centered at $\langle E \rangle-\langle \Psi(T)|\hat{H}_{I}|\Psi(T)\rangle$ (see Figure \ref{App_A}). For the width of this window, we tried four different values ranging from 0.1 to 0.0125 hartree, and found that they produced the same results with a good approximation.

\bigskip

\bibliographystyle{apsrev4-1}
\bibliography{bibliography}

\begin{thebibliography}{53}%
\makeatletter
\providecommand \@ifxundefined [1]{%
 \@ifx{#1\undefined}
}%
\providecommand \@ifnum [1]{%
 \ifnum #1\expandafter \@firstoftwo
 \else \expandafter \@secondoftwo
 \fi
}%
\providecommand \@ifx [1]{%
 \ifx #1\expandafter \@firstoftwo
 \else \expandafter \@secondoftwo
 \fi
}%
\providecommand \natexlab [1]{#1}%
\providecommand \enquote  [1]{``#1''}%
\providecommand \bibnamefont  [1]{#1}%
\providecommand \bibfnamefont [1]{#1}%
\providecommand \citenamefont [1]{#1}%
\providecommand \href@noop [0]{\@secondoftwo}%
\providecommand \href [0]{\begingroup \@sanitize@url \@href}%
\providecommand \@href[1]{\@@startlink{#1}\@@href}%
\providecommand \@@href[1]{\endgroup#1\@@endlink}%
\providecommand \@sanitize@url [0]{\catcode `\\12\catcode `\$12\catcode
  `\&12\catcode `\#12\catcode `\^12\catcode `\_12\catcode `\%12\relax}%
\providecommand \@@startlink[1]{}%
\providecommand \@@endlink[0]{}%
\providecommand \url  [0]{\begingroup\@sanitize@url \@url }%
\providecommand \@url [1]{\endgroup\@href {#1}{\urlprefix }}%
\providecommand \urlprefix  [0]{URL }%
\providecommand \Eprint [0]{\href }%
\providecommand \doibase [0]{http://dx.doi.org/}%
\providecommand \selectlanguage [0]{\@gobble}%
\providecommand \bibinfo  [0]{\@secondoftwo}%
\providecommand \bibfield  [0]{\@secondoftwo}%
\providecommand \translation [1]{[#1]}%
\providecommand \BibitemOpen [0]{}%
\providecommand \bibitemStop [0]{}%
\providecommand \bibitemNoStop [0]{.\EOS\space}%
\providecommand \EOS [0]{\spacefactor3000\relax}%
\providecommand \BibitemShut  [1]{\csname bibitem#1\endcsname}%
\let\auto@bib@innerbib\@empty
%</preamble>
\bibitem [{\citenamefont {Nandkishore}\ and\ \citenamefont
  {Huse}(2015)}]{Huse}%
  \BibitemOpen
  \bibfield  {author} {\bibinfo {author} {\bibfnamefont {R.}~\bibnamefont
  {Nandkishore}}\ and\ \bibinfo {author} {\bibfnamefont {D.~A.}\ \bibnamefont
  {Huse}},\ }\href@noop {} {\bibfield  {journal} {\bibinfo  {journal} {Annu.
  Rev. Condens. Matter Phys.}\ }\textbf {\bibinfo {volume} {6}},\ \bibinfo
  {pages} {15} (\bibinfo {year} {2015})}\BibitemShut {NoStop}%
\bibitem [{\citenamefont {Gogolin}\ and\ \citenamefont
  {Eisert}(2016)}]{Gogolin}%
  \BibitemOpen
  \bibfield  {author} {\bibinfo {author} {\bibfnamefont {C.}~\bibnamefont
  {Gogolin}}\ and\ \bibinfo {author} {\bibfnamefont {J.}~\bibnamefont
  {Eisert}},\ }\href@noop {} {\bibfield  {journal} {\bibinfo  {journal} {Rep.
  Prog. Phys.}\ }\textbf {\bibinfo {volume} {79}},\ \bibinfo {pages} {1}
  (\bibinfo {year} {2016})}\BibitemShut {NoStop}%
\bibitem [{\citenamefont {D'Alessio}\ \emph {et~al.}(2016)\citenamefont
  {D'Alessio}, \citenamefont {Kafri}, \citenamefont {Polkovnikov},\ and\
  \citenamefont {Rigol}}]{Dalessio}%
  \BibitemOpen
  \bibfield  {author} {\bibinfo {author} {\bibfnamefont {L.}~\bibnamefont
  {D'Alessio}}, \bibinfo {author} {\bibfnamefont {Y.}~\bibnamefont {Kafri}},
  \bibinfo {author} {\bibfnamefont {A.}~\bibnamefont {Polkovnikov}}, \ and\
  \bibinfo {author} {\bibfnamefont {M.}~\bibnamefont {Rigol}},\ }\href@noop {}
  {\bibfield  {journal} {\bibinfo  {journal} {Adv. Phys.}\ }\textbf {\bibinfo
  {volume} {65}},\ \bibinfo {pages} {239} (\bibinfo {year} {2016})}\BibitemShut
  {NoStop}%
\bibitem [{\citenamefont {von Neumann}(1929)}]{Neumann}%
  \BibitemOpen
  \bibfield  {author} {\bibinfo {author} {\bibfnamefont {J.}~\bibnamefont {von
  Neumann}},\ }\href@noop {} {\bibfield  {journal} {\bibinfo  {journal} {Z.
  Phys.}\ }\textbf {\bibinfo {volume} {57}},\ \bibinfo {pages} {30} (\bibinfo
  {year} {1929})},\ \bibinfo {note} {translated to English in Eur. Phys. J. H
  \textbf{35}, 201 (2010).}\BibitemShut {Stop}%
\bibitem [{\citenamefont {Uzer}(1991)}]{Uzer}%
  \BibitemOpen
  \bibfield  {author} {\bibinfo {author} {\bibfnamefont {T.}~\bibnamefont
  {Uzer}},\ }\href@noop {} {\bibfield  {journal} {\bibinfo  {journal} {Phys.
  Rep.}\ }\textbf {\bibinfo {volume} {199}},\ \bibinfo {pages} {73} (\bibinfo
  {year} {1991})}\BibitemShut {NoStop}%
\bibitem [{\citenamefont {Nesbitt}\ and\ \citenamefont {Field}(1996)}]{Field}%
  \BibitemOpen
  \bibfield  {author} {\bibinfo {author} {\bibfnamefont {D.~J.}\ \bibnamefont
  {Nesbitt}}\ and\ \bibinfo {author} {\bibfnamefont {R.~W.}\ \bibnamefont
  {Field}},\ }\href@noop {} {\bibfield  {journal} {\bibinfo  {journal} {J.
  Phys. Chem.}\ }\textbf {\bibinfo {volume} {100}},\ \bibinfo {pages} {12735}
  (\bibinfo {year} {1996})}\BibitemShut {NoStop}%
\bibitem [{\citenamefont {Marcus}(1952)}]{Marcus}%
  \BibitemOpen
  \bibfield  {author} {\bibinfo {author} {\bibfnamefont {R.~A.}\ \bibnamefont
  {Marcus}},\ }\href@noop {} {\bibfield  {journal} {\bibinfo  {journal} {J.
  Chem. Phys.}\ }\textbf {\bibinfo {volume} {20}},\ \bibinfo {pages} {359}
  (\bibinfo {year} {1952})}\BibitemShut {NoStop}%
\bibitem [{\citenamefont {Baer}\ and\ \citenamefont {Hase}(1996)}]{Baer-Hase}%
  \BibitemOpen
  \bibfield  {author} {\bibinfo {author} {\bibfnamefont {T.}~\bibnamefont
  {Baer}}\ and\ \bibinfo {author} {\bibfnamefont {W.}~\bibnamefont {Hase}},\
  }\href {https://books.google.com.co/books?id=zL2GAAAAIAAJ} {\emph {\bibinfo
  {title} {Unimolecular Reaction Dynamics}}}\ (\bibinfo  {publisher} {Oxford
  University Press, Oxford},\ \bibinfo {year} {1996})\BibitemShut {NoStop}%
\bibitem [{\citenamefont {Logan}\ and\ \citenamefont
  {Wolynes}(1990)}]{Wolynes1}%
  \BibitemOpen
  \bibfield  {author} {\bibinfo {author} {\bibfnamefont {D.~E.}\ \bibnamefont
  {Logan}}\ and\ \bibinfo {author} {\bibfnamefont {P.~G.}\ \bibnamefont
  {Wolynes}},\ }\href@noop {} {\bibfield  {journal} {\bibinfo  {journal} {J.
  Chem. Phys.}\ }\textbf {\bibinfo {volume} {93}},\ \bibinfo {pages} {4994}
  (\bibinfo {year} {1990})}\BibitemShut {NoStop}%
\bibitem [{\citenamefont {Bunker}\ and\ \citenamefont {Hase}(1973)}]{Bunker73}%
  \BibitemOpen
  \bibfield  {author} {\bibinfo {author} {\bibfnamefont {D.~L.}\ \bibnamefont
  {Bunker}}\ and\ \bibinfo {author} {\bibfnamefont {W.~L.}\ \bibnamefont
  {Hase}},\ }\href@noop {} {\bibfield  {journal} {\bibinfo  {journal} {J. Chem.
  Phys.}\ }\textbf {\bibinfo {volume} {59}},\ \bibinfo {pages} {4621} (\bibinfo
  {year} {1973})}\BibitemShut {NoStop}%
\bibitem [{\citenamefont {Marcus}\ \emph {et~al.}(1984)\citenamefont {Marcus},
  \citenamefont {Hase},\ and\ \citenamefont {Swamy}}]{Marcus84}%
  \BibitemOpen
  \bibfield  {author} {\bibinfo {author} {\bibfnamefont {R.~A.}\ \bibnamefont
  {Marcus}}, \bibinfo {author} {\bibfnamefont {W.~L.}\ \bibnamefont {Hase}}, \
  and\ \bibinfo {author} {\bibfnamefont {K.}~\bibnamefont {Swamy}},\
  }\href@noop {} {\bibfield  {journal} {\bibinfo  {journal} {J. Phys. Chem.}\
  }\textbf {\bibinfo {volume} {88}},\ \bibinfo {pages} {6717} (\bibinfo {year}
  {1984})}\BibitemShut {NoStop}%
\bibitem [{\citenamefont {Rice}\ \emph {et~al.}(1990)\citenamefont {Rice},
  \citenamefont {Chung},\ and\ \citenamefont {Baronavski}}]{Rice90}%
  \BibitemOpen
  \bibfield  {author} {\bibinfo {author} {\bibfnamefont {J.~K.}\ \bibnamefont
  {Rice}}, \bibinfo {author} {\bibfnamefont {Y.~C.}\ \bibnamefont {Chung}}, \
  and\ \bibinfo {author} {\bibfnamefont {A.}~\bibnamefont {Baronavski}},\
  }\href@noop {} {\bibfield  {journal} {\bibinfo  {journal} {Chem. Phys.
  Lett.}\ }\textbf {\bibinfo {volume} {167}},\ \bibinfo {pages} {151} (\bibinfo
  {year} {1990})}\BibitemShut {NoStop}%
\bibitem [{\citenamefont {Armenise}\ \emph {et~al.}(1992)\citenamefont
  {Armenise}, \citenamefont {Capitelli}, \citenamefont {Garcia}, \citenamefont
  {Gorse}, \citenamefont {Lagana},\ and\ \citenamefont {Longo}}]{Armenise92}%
  \BibitemOpen
  \bibfield  {author} {\bibinfo {author} {\bibfnamefont {I.}~\bibnamefont
  {Armenise}}, \bibinfo {author} {\bibfnamefont {M.}~\bibnamefont {Capitelli}},
  \bibinfo {author} {\bibfnamefont {E.}~\bibnamefont {Garcia}}, \bibinfo
  {author} {\bibfnamefont {C.}~\bibnamefont {Gorse}}, \bibinfo {author}
  {\bibfnamefont {A.}~\bibnamefont {Lagana}}, \ and\ \bibinfo {author}
  {\bibfnamefont {S.}~\bibnamefont {Longo}},\ }\href@noop {} {\bibfield
  {journal} {\bibinfo  {journal} {Chem. Phys. Lett.}\ }\textbf {\bibinfo
  {volume} {200}},\ \bibinfo {pages} {597} (\bibinfo {year}
  {1992})}\BibitemShut {NoStop}%
\bibitem [{\citenamefont {Tonner}\ and\ \citenamefont
  {McMahon}(2000)}]{Tonner2000}%
  \BibitemOpen
  \bibfield  {author} {\bibinfo {author} {\bibfnamefont {D.~S.}\ \bibnamefont
  {Tonner}}\ and\ \bibinfo {author} {\bibfnamefont {T.~B.}\ \bibnamefont
  {McMahon}},\ }\href@noop {} {\bibfield  {journal} {\bibinfo  {journal} {J.
  Am. Chem. Soc.}\ }\textbf {\bibinfo {volume} {122}},\ \bibinfo {pages} {8783}
  (\bibinfo {year} {2000})}\BibitemShut {NoStop}%
\bibitem [{\citenamefont {Utz}(2009)}]{Utz2009}%
  \BibitemOpen
  \bibfield  {author} {\bibinfo {author} {\bibfnamefont {A.~L.}\ \bibnamefont
  {Utz}},\ }\href@noop {} {\bibfield  {journal} {\bibinfo  {journal} {Curr.
  Opin. Solid State Mater. Sci.}\ }\textbf {\bibinfo {volume} {13}},\ \bibinfo
  {pages} {4} (\bibinfo {year} {2009})}\BibitemShut {NoStop}%
\bibitem [{\citenamefont {Gruebele}\ and\ \citenamefont
  {Wolynes}(2004)}]{Gruebele2}%
  \BibitemOpen
  \bibfield  {author} {\bibinfo {author} {\bibfnamefont {M.}~\bibnamefont
  {Gruebele}}\ and\ \bibinfo {author} {\bibfnamefont {P.~G.}\ \bibnamefont
  {Wolynes}},\ }\href@noop {} {\bibfield  {journal} {\bibinfo  {journal} {Acc.
  Chem. Res.}\ }\textbf {\bibinfo {volume} {37}},\ \bibinfo {pages} {261}
  (\bibinfo {year} {2004})}\BibitemShut {NoStop}%
\bibitem [{\citenamefont {Kaufman}\ \emph {et~al.}(2016)\citenamefont
  {Kaufman}, \citenamefont {Tai}, \citenamefont {Lukin}, \citenamefont
  {Rispoli}, \citenamefont {Schittko}, \citenamefont {Preiss},\ and\
  \citenamefont {Greiner}}]{Kaufman}%
  \BibitemOpen
  \bibfield  {author} {\bibinfo {author} {\bibfnamefont {A.~M.}\ \bibnamefont
  {Kaufman}}, \bibinfo {author} {\bibfnamefont {M.~E.}\ \bibnamefont {Tai}},
  \bibinfo {author} {\bibfnamefont {A.}~\bibnamefont {Lukin}}, \bibinfo
  {author} {\bibfnamefont {M.}~\bibnamefont {Rispoli}}, \bibinfo {author}
  {\bibfnamefont {R.}~\bibnamefont {Schittko}}, \bibinfo {author}
  {\bibfnamefont {P.}~\bibnamefont {Preiss}}, \ and\ \bibinfo {author}
  {\bibfnamefont {M.}~\bibnamefont {Greiner}},\ }\href@noop {} {\bibfield
  {journal} {\bibinfo  {journal} {Science}\ }\textbf {\bibinfo {volume}
  {353}},\ \bibinfo {pages} {794} (\bibinfo {year} {2016})}\BibitemShut
  {NoStop}%
\bibitem [{\citenamefont {Neill}\ \emph {et~al.}(2016)\citenamefont {Neill},
  \citenamefont {Roushan}, \citenamefont {Fang}, \citenamefont {Chen},
  \citenamefont {Kolodrubetz}, \citenamefont {Chen}, \citenamefont {Megrant},
  \citenamefont {Barends}, \citenamefont {Campbell}, \citenamefont {Chiaro},
  \citenamefont {Dunsworth}, \citenamefont {Jeffrey}, \citenamefont {Kelly},
  \citenamefont {Mutus}, \citenamefont {O'Malley}, \citenamefont {Quintana},
  \citenamefont {Sank}, \citenamefont {Vainsencher}, \citenamefont {Wenner},
  \citenamefont {White}, \citenamefont {Polkovnikov},\ and\ \citenamefont
  {Martinis}}]{Neill}%
  \BibitemOpen
  \bibfield  {author} {\bibinfo {author} {\bibfnamefont {C.}~\bibnamefont
  {Neill}}, \bibinfo {author} {\bibfnamefont {P.}~\bibnamefont {Roushan}},
  \bibinfo {author} {\bibfnamefont {M.}~\bibnamefont {Fang}}, \bibinfo {author}
  {\bibfnamefont {Y.}~\bibnamefont {Chen}}, \bibinfo {author} {\bibfnamefont
  {M.}~\bibnamefont {Kolodrubetz}}, \bibinfo {author} {\bibfnamefont
  {Z.}~\bibnamefont {Chen}}, \bibinfo {author} {\bibfnamefont {A.}~\bibnamefont
  {Megrant}}, \bibinfo {author} {\bibfnamefont {R.}~\bibnamefont {Barends}},
  \bibinfo {author} {\bibfnamefont {B.}~\bibnamefont {Campbell}}, \bibinfo
  {author} {\bibfnamefont {B.}~\bibnamefont {Chiaro}}, \bibinfo {author}
  {\bibfnamefont {A.}~\bibnamefont {Dunsworth}}, \bibinfo {author}
  {\bibfnamefont {E.}~\bibnamefont {Jeffrey}}, \bibinfo {author} {\bibfnamefont
  {J.}~\bibnamefont {Kelly}}, \bibinfo {author} {\bibfnamefont
  {J.}~\bibnamefont {Mutus}}, \bibinfo {author} {\bibfnamefont {P.~J.~J.}\
  \bibnamefont {O'Malley}}, \bibinfo {author} {\bibfnamefont {C.}~\bibnamefont
  {Quintana}}, \bibinfo {author} {\bibfnamefont {D.}~\bibnamefont {Sank}},
  \bibinfo {author} {\bibfnamefont {A.}~\bibnamefont {Vainsencher}}, \bibinfo
  {author} {\bibfnamefont {J.}~\bibnamefont {Wenner}}, \bibinfo {author}
  {\bibfnamefont {T.~C.}\ \bibnamefont {White}}, \bibinfo {author}
  {\bibfnamefont {A.}~\bibnamefont {Polkovnikov}}, \ and\ \bibinfo {author}
  {\bibfnamefont {J.~M.}\ \bibnamefont {Martinis}},\ }\href@noop {} {\bibfield
  {journal} {\bibinfo  {journal} {Nat. Phys.}\ }\textbf {\bibinfo {volume}
  {12}},\ \bibinfo {pages} {1037} (\bibinfo {year} {2016})}\BibitemShut
  {NoStop}%
\bibitem [{\citenamefont {Santos}\ \emph {et~al.}(2011)\citenamefont {Santos},
  \citenamefont {Polkovnikov},\ and\ \citenamefont {Rigol}}]{Santos}%
  \BibitemOpen
  \bibfield  {author} {\bibinfo {author} {\bibfnamefont {L.~F.}\ \bibnamefont
  {Santos}}, \bibinfo {author} {\bibfnamefont {A.}~\bibnamefont {Polkovnikov}},
  \ and\ \bibinfo {author} {\bibfnamefont {M.}~\bibnamefont {Rigol}},\
  }\href@noop {} {\bibfield  {journal} {\bibinfo  {journal} {Phys. Rev. Lett.}\
  }\textbf {\bibinfo {volume} {107}},\ \bibinfo {pages} {040601} (\bibinfo
  {year} {2011})}\BibitemShut {NoStop}%
\bibitem [{\citenamefont {Shapiro}\ and\ \citenamefont
  {Brumer}(2012)}]{Brumershapiro2}%
  \BibitemOpen
  \bibfield  {author} {\bibinfo {author} {\bibfnamefont {M.}~\bibnamefont
  {Shapiro}}\ and\ \bibinfo {author} {\bibfnamefont {P.}~\bibnamefont
  {Brumer}},\ }\href@noop {} {\emph {\bibinfo {title} {Quantum Control of
  Molecular Processes}}},\ \bibinfo {edition} {2nd}\ ed.\ (\bibinfo
  {publisher} {WILEY-VCH Verlag GmbH \& Co., Weinheim},\ \bibinfo {year}
  {2012})\BibitemShut {NoStop}%
\bibitem [{\citenamefont {Bixon}\ and\ \citenamefont
  {Jortner}(1968)}]{Jortner}%
  \BibitemOpen
  \bibfield  {author} {\bibinfo {author} {\bibfnamefont {M.}~\bibnamefont
  {Bixon}}\ and\ \bibinfo {author} {\bibfnamefont {J.}~\bibnamefont
  {Jortner}},\ }\href@noop {} {\bibfield  {journal} {\bibinfo  {journal} {J.
  Chem. Phys.}\ }\textbf {\bibinfo {volume} {48}},\ \bibinfo {pages} {715}
  (\bibinfo {year} {1968})}\BibitemShut {NoStop}%
\bibitem [{\citenamefont {Stewart}\ and\ \citenamefont
  {McDonald}(1983)}]{McDonald1}%
  \BibitemOpen
  \bibfield  {author} {\bibinfo {author} {\bibfnamefont {G.~M.}\ \bibnamefont
  {Stewart}}\ and\ \bibinfo {author} {\bibfnamefont {J.~D.}\ \bibnamefont
  {McDonald}},\ }\href@noop {} {\bibfield  {journal} {\bibinfo  {journal} {J.
  Chem. Phys.}\ }\textbf {\bibinfo {volume} {78}},\ \bibinfo {pages} {3907}
  (\bibinfo {year} {1983})}\BibitemShut {NoStop}%
\bibitem [{\citenamefont {Hutchinson}\ \emph {et~al.}(1984)\citenamefont
  {Hutchinson}, \citenamefont {Sibert},\ and\ \citenamefont {Hynes}}]{sibert}%
  \BibitemOpen
  \bibfield  {author} {\bibinfo {author} {\bibfnamefont {J.~S.}\ \bibnamefont
  {Hutchinson}}, \bibinfo {author} {\bibfnamefont {E.~L.}\ \bibnamefont
  {Sibert}}, \ and\ \bibinfo {author} {\bibfnamefont {J.~T.}\ \bibnamefont
  {Hynes}},\ }\href@noop {} {\bibfield  {journal} {\bibinfo  {journal} {J.
  Chem. Phys.}\ }\textbf {\bibinfo {volume} {81}},\ \bibinfo {pages} {1314}
  (\bibinfo {year} {1984})}\BibitemShut {NoStop}%
\bibitem [{\citenamefont {Stuchebrukhov}\ and\ \citenamefont
  {Marcus}(1993)}]{Stuchebrukhov1}%
  \BibitemOpen
  \bibfield  {author} {\bibinfo {author} {\bibfnamefont {A.~A.}\ \bibnamefont
  {Stuchebrukhov}}\ and\ \bibinfo {author} {\bibfnamefont {R.~A.}\ \bibnamefont
  {Marcus}},\ }\href@noop {} {\bibfield  {journal} {\bibinfo  {journal} {J.
  Chem. Phys.}\ }\textbf {\bibinfo {volume} {98}},\ \bibinfo {pages} {6044}
  (\bibinfo {year} {1993})}\BibitemShut {NoStop}%
\bibitem [{\citenamefont {Stuchebrukhov}\ \emph {et~al.}(1993)\citenamefont
  {Stuchebrukhov}, \citenamefont {Mehta},\ and\ \citenamefont
  {Marcus}}]{Stuchebrukhov2}%
  \BibitemOpen
  \bibfield  {author} {\bibinfo {author} {\bibfnamefont {A.~A.}\ \bibnamefont
  {Stuchebrukhov}}, \bibinfo {author} {\bibfnamefont {A.}~\bibnamefont
  {Mehta}}, \ and\ \bibinfo {author} {\bibfnamefont {R.~A.}\ \bibnamefont
  {Marcus}},\ }\href@noop {} {\bibfield  {journal} {\bibinfo  {journal} {J.
  Phys. Chem.}\ }\textbf {\bibinfo {volume} {97}},\ \bibinfo {pages} {12491}
  (\bibinfo {year} {1993})}\BibitemShut {NoStop}%
\bibitem [{\citenamefont {Gruebele}(2000)}]{Gruebele1}%
  \BibitemOpen
  \bibfield  {author} {\bibinfo {author} {\bibfnamefont {M.}~\bibnamefont
  {Gruebele}},\ }\href@noop {} {\bibfield  {journal} {\bibinfo  {journal} {Adv.
  Chem. Phys.}\ }\textbf {\bibinfo {volume} {114}},\ \bibinfo {pages} {193}
  (\bibinfo {year} {2000})}\BibitemShut {NoStop}%
\bibitem [{\citenamefont {Leitner}(2015)}]{leitner}%
  \BibitemOpen
  \bibfield  {author} {\bibinfo {author} {\bibfnamefont {D.~M.}\ \bibnamefont
  {Leitner}},\ }\href@noop {} {\bibfield  {journal} {\bibinfo  {journal} {Adv.
  Phys.}\ }\textbf {\bibinfo {volume} {64}},\ \bibinfo {pages} {445} (\bibinfo
  {year} {2015})}\BibitemShut {NoStop}%
\bibitem [{\citenamefont {Carter}\ and\ \citenamefont
  {Brumer}(1982)}]{CarterBrumer}%
  \BibitemOpen
  \bibfield  {author} {\bibinfo {author} {\bibfnamefont {D.}~\bibnamefont
  {Carter}}\ and\ \bibinfo {author} {\bibfnamefont {P.}~\bibnamefont
  {Brumer}},\ }\href@noop {} {\bibfield  {journal} {\bibinfo  {journal} {J.
  Chem. Phys.}\ }\textbf {\bibinfo {volume} {77}},\ \bibinfo {pages} {4208}
  (\bibinfo {year} {1982})}\BibitemShut {NoStop}%
\bibitem [{\citenamefont {Carter}\ and\ \citenamefont
  {Brumer}(1983)}]{carter83}%
  \BibitemOpen
  \bibfield  {author} {\bibinfo {author} {\bibfnamefont {D.}~\bibnamefont
  {Carter}}\ and\ \bibinfo {author} {\bibfnamefont {P.}~\bibnamefont
  {Brumer}},\ }\href@noop {} {\bibfield  {journal} {\bibinfo  {journal} {J.
  Chem. Phys.}\ }\textbf {\bibinfo {volume} {78}},\ \bibinfo {pages} {2104
  (Erratum)} (\bibinfo {year} {1983})}\BibitemShut {NoStop}%
\bibitem [{\citenamefont {Davis}\ and\ \citenamefont
  {Wagner}(1984)}]{DavisWagner}%
  \BibitemOpen
  \bibfield  {author} {\bibinfo {author} {\bibfnamefont {M.~J.}\ \bibnamefont
  {Davis}}\ and\ \bibinfo {author} {\bibfnamefont {A.~F.}\ \bibnamefont
  {Wagner}},\ }in\ \href@noop {} {\emph {\bibinfo {booktitle} {Resonances in
  Electron-Molecule Scattering, van der Waals Complexes, and Reactive Chemical
  Dynamics}}},\ \bibinfo {editor} {edited by\ \bibinfo {editor} {\bibfnamefont
  {D.~G.}\ \bibnamefont {Truhlar}}}\ (\bibinfo  {publisher} {American Chemical
  Society, Washington, D.C.},\ \bibinfo {year} {1984})\ Chap.~\bibinfo
  {chapter} {18}, pp.\ \bibinfo {pages} {337--349}\BibitemShut {NoStop}%
\bibitem [{\citenamefont {Davis}(1984)}]{davis84}%
  \BibitemOpen
  \bibfield  {author} {\bibinfo {author} {\bibfnamefont {M.~J.}\ \bibnamefont
  {Davis}},\ }\href@noop {} {\bibfield  {journal} {\bibinfo  {journal} {Chem.
  Phys. Lett.}\ }\textbf {\bibinfo {volume} {110}},\ \bibinfo {pages} {491}
  (\bibinfo {year} {1984})}\BibitemShut {NoStop}%
\bibitem [{\citenamefont {Davis}(1985)}]{Davis}%
  \BibitemOpen
  \bibfield  {author} {\bibinfo {author} {\bibfnamefont {M.~J.}\ \bibnamefont
  {Davis}},\ }\href@noop {} {\bibfield  {journal} {\bibinfo  {journal} {J.
  Chem. Phys.}\ }\textbf {\bibinfo {volume} {83}},\ \bibinfo {pages} {1016}
  (\bibinfo {year} {1985})}\BibitemShut {NoStop}%
\bibitem [{\citenamefont {Martens}\ \emph {et~al.}(1987)\citenamefont
  {Martens}, \citenamefont {Davis},\ and\ \citenamefont {Ezra}}]{Martens}%
  \BibitemOpen
  \bibfield  {author} {\bibinfo {author} {\bibfnamefont {C.~C.}\ \bibnamefont
  {Martens}}, \bibinfo {author} {\bibfnamefont {M.~J.}\ \bibnamefont {Davis}},
  \ and\ \bibinfo {author} {\bibfnamefont {G.~S.}\ \bibnamefont {Ezra}},\
  }\href@noop {} {\bibfield  {journal} {\bibinfo  {journal} {Chem. Phys.
  Lett.}\ }\textbf {\bibinfo {volume} {142}},\ \bibinfo {pages} {519} (\bibinfo
  {year} {1987})}\BibitemShut {NoStop}%
\bibitem [{\citenamefont {Shchekinova}\ \emph {et~al.}(2004)\citenamefont
  {Shchekinova}, \citenamefont {Chandre}, \citenamefont {Lan},\ and\
  \citenamefont {Uzer}}]{Chandre}%
  \BibitemOpen
  \bibfield  {author} {\bibinfo {author} {\bibfnamefont {E.}~\bibnamefont
  {Shchekinova}}, \bibinfo {author} {\bibfnamefont {C.}~\bibnamefont
  {Chandre}}, \bibinfo {author} {\bibfnamefont {Y.}~\bibnamefont {Lan}}, \ and\
  \bibinfo {author} {\bibfnamefont {T.}~\bibnamefont {Uzer}},\ }\href@noop {}
  {\bibfield  {journal} {\bibinfo  {journal} {J. Chem. Phys.}\ }\textbf
  {\bibinfo {volume} {121}},\ \bibinfo {pages} {3471} (\bibinfo {year}
  {2004})}\BibitemShut {NoStop}%
\bibitem [{\citenamefont {Pa\ifmmode~\check{s}\else \v{s}\fi{}kauskas}\ \emph
  {et~al.}(2008)\citenamefont {Pa\ifmmode~\check{s}\else \v{s}\fi{}kauskas},
  \citenamefont {Chandre},\ and\ \citenamefont {Uzer}}]{PaskaukasPRL}%
  \BibitemOpen
  \bibfield  {author} {\bibinfo {author} {\bibfnamefont {R.}~\bibnamefont
  {Pa\ifmmode~\check{s}\else \v{s}\fi{}kauskas}}, \bibinfo {author}
  {\bibfnamefont {C.}~\bibnamefont {Chandre}}, \ and\ \bibinfo {author}
  {\bibfnamefont {T.}~\bibnamefont {Uzer}},\ }\href@noop {} {\bibfield
  {journal} {\bibinfo  {journal} {Phys. Rev. Lett.}\ }\textbf {\bibinfo
  {volume} {100}},\ \bibinfo {pages} {083001} (\bibinfo {year}
  {2008})}\BibitemShut {NoStop}%
\bibitem [{\citenamefont {Pa\ifmmode~\check{s}\else \v{s}\fi{}kauskas}\ \emph
  {et~al.}(2009)\citenamefont {Pa\ifmmode~\check{s}\else \v{s}\fi{}kauskas},
  \citenamefont {Chandre},\ and\ \citenamefont {Uzer}}]{paskaukas}%
  \BibitemOpen
  \bibfield  {author} {\bibinfo {author} {\bibfnamefont {R.}~\bibnamefont
  {Pa\ifmmode~\check{s}\else \v{s}\fi{}kauskas}}, \bibinfo {author}
  {\bibfnamefont {C.}~\bibnamefont {Chandre}}, \ and\ \bibinfo {author}
  {\bibfnamefont {T.}~\bibnamefont {Uzer}},\ }\href@noop {} {\bibfield
  {journal} {\bibinfo  {journal} {J. Chem. Phys.}\ }\textbf {\bibinfo {volume}
  {130}},\ \bibinfo {pages} {164105} (\bibinfo {year} {2009})}\BibitemShut
  {NoStop}%
\bibitem [{\citenamefont {Gibson}\ \emph {et~al.}(1987)\citenamefont {Gibson},
  \citenamefont {Schatz}, \citenamefont {Ratner},\ and\ \citenamefont
  {Davis}}]{Gibson}%
  \BibitemOpen
  \bibfield  {author} {\bibinfo {author} {\bibfnamefont {L.~L.}\ \bibnamefont
  {Gibson}}, \bibinfo {author} {\bibfnamefont {G.~C.}\ \bibnamefont {Schatz}},
  \bibinfo {author} {\bibfnamefont {M.~A.}\ \bibnamefont {Ratner}}, \ and\
  \bibinfo {author} {\bibfnamefont {M.~J.}\ \bibnamefont {Davis}},\ }\href@noop
  {} {\bibfield  {journal} {\bibinfo  {journal} {J. Chem. Phys.}\ }\textbf
  {\bibinfo {volume} {86}},\ \bibinfo {pages} {3263} (\bibinfo {year}
  {1987})}\BibitemShut {NoStop}%
\bibitem [{\citenamefont {Xie}\ \emph {et~al.}(2001)\citenamefont {Xie},
  \citenamefont {Lu}, \citenamefont {Xu},\ and\ \citenamefont {Yan}}]{Xie}%
  \BibitemOpen
  \bibfield  {author} {\bibinfo {author} {\bibfnamefont {D.}~\bibnamefont
  {Xie}}, \bibinfo {author} {\bibfnamefont {Y.}~\bibnamefont {Lu}}, \bibinfo
  {author} {\bibfnamefont {D.}~\bibnamefont {Xu}}, \ and\ \bibinfo {author}
  {\bibfnamefont {G.}~\bibnamefont {Yan}},\ }\href@noop {} {\bibfield
  {journal} {\bibinfo  {journal} {Chem. Phys.}\ }\textbf {\bibinfo {volume}
  {270}},\ \bibinfo {pages} {415} (\bibinfo {year} {2001})}\BibitemShut
  {NoStop}%
\bibitem [{\citenamefont {Beck}\ \emph {et~al.}(2000)\citenamefont {Beck},
  \citenamefont {J{\"a}ckle}, \citenamefont {Worth},\ and\ \citenamefont
  {Meyer}}]{MCTDHPR}%
  \BibitemOpen
  \bibfield  {author} {\bibinfo {author} {\bibfnamefont {M.~H.}\ \bibnamefont
  {Beck}}, \bibinfo {author} {\bibfnamefont {A.}~\bibnamefont {J{\"a}ckle}},
  \bibinfo {author} {\bibfnamefont {G.~A.}\ \bibnamefont {Worth}}, \ and\
  \bibinfo {author} {\bibfnamefont {H.-D.}\ \bibnamefont {Meyer}},\ }\href@noop
  {} {\bibfield  {journal} {\bibinfo  {journal} {Phys. Rep.}\ }\textbf
  {\bibinfo {volume} {324}},\ \bibinfo {pages} {1} (\bibinfo {year}
  {2000})}\BibitemShut {NoStop}%
\bibitem [{\citenamefont {Flambaum}\ and\ \citenamefont
  {Izrailev}(1997)}]{Flambaum1997}%
  \BibitemOpen
  \bibfield  {author} {\bibinfo {author} {\bibfnamefont {V.~V.}\ \bibnamefont
  {Flambaum}}\ and\ \bibinfo {author} {\bibfnamefont {F.~M.}\ \bibnamefont
  {Izrailev}},\ }\href@noop {} {\bibfield  {journal} {\bibinfo  {journal}
  {Phys. Rev. E}\ }\textbf {\bibinfo {volume} {56}},\ \bibinfo {pages} {5144}
  (\bibinfo {year} {1997})}\BibitemShut {NoStop}%
\bibitem [{\citenamefont {Beck}\ and\ \citenamefont {Meyer}(2001)}]{beck}%
  \BibitemOpen
  \bibfield  {author} {\bibinfo {author} {\bibfnamefont {M.}~\bibnamefont
  {Beck}}\ and\ \bibinfo {author} {\bibfnamefont {H.-D.}\ \bibnamefont
  {Meyer}},\ }\href@noop {} {\bibfield  {journal} {\bibinfo  {journal} {J.
  Chem. Phys.}\ }\textbf {\bibinfo {volume} {114}},\ \bibinfo {pages} {2036}
  (\bibinfo {year} {2001})}\BibitemShut {NoStop}%
\bibitem [{\citenamefont {Gerbasi}\ \emph {et~al.}(2007)\citenamefont
  {Gerbasi}, \citenamefont {Sanz}, \citenamefont {Christopher}, \citenamefont
  {Shapiro},\ and\ \citenamefont {Brumer}}]{Gerbasi}%
  \BibitemOpen
  \bibfield  {author} {\bibinfo {author} {\bibfnamefont {D.}~\bibnamefont
  {Gerbasi}}, \bibinfo {author} {\bibfnamefont {A.~S.}\ \bibnamefont {Sanz}},
  \bibinfo {author} {\bibfnamefont {P.~S.}\ \bibnamefont {Christopher}},
  \bibinfo {author} {\bibfnamefont {M.}~\bibnamefont {Shapiro}}, \ and\
  \bibinfo {author} {\bibfnamefont {P.}~\bibnamefont {Brumer}},\ }\href@noop {}
  {\bibfield  {journal} {\bibinfo  {journal} {J. Chem. Phys.}\ }\textbf
  {\bibinfo {volume} {126}},\ \bibinfo {eid} {124307} (\bibinfo {year}
  {2007})}\BibitemShut {NoStop}%
\bibitem [{\citenamefont {Meyer}\ \emph {et~al.}(2009)\citenamefont {Meyer},
  \citenamefont {Gatti},\ and\ \citenamefont {Worth}}]{meyer2009}%
  \BibitemOpen
  \bibinfo {editor} {\bibfnamefont {H.~D.}\ \bibnamefont {Meyer}}, \bibinfo
  {editor} {\bibfnamefont {F.}~\bibnamefont {Gatti}}, \ and\ \bibinfo {editor}
  {\bibfnamefont {G.~A.}\ \bibnamefont {Worth}},\ eds.,\ \href@noop {} {\emph
  {\bibinfo {title} {Multidimensional Quantum Dynamics: {MCTDH} Theory and
  Applications}}}\ (\bibinfo  {publisher} {John Wiley \& Sons, Weinheim},\
  \bibinfo {year} {2009})\BibitemShut {NoStop}%
\bibitem [{\citenamefont {Worth}\ \emph {et~al.}(2000)\citenamefont {Worth},
  \citenamefont {Beck}, \citenamefont {J{\"a}ckle},\ and\ \citenamefont
  {Meyer}}]{paquete}%
  \BibitemOpen
  \bibfield  {author} {\bibinfo {author} {\bibfnamefont {G.~A.}\ \bibnamefont
  {Worth}}, \bibinfo {author} {\bibfnamefont {M.~H.}\ \bibnamefont {Beck}},
  \bibinfo {author} {\bibfnamefont {A.}~\bibnamefont {J{\"a}ckle}}, \ and\
  \bibinfo {author} {\bibfnamefont {H.-D.}\ \bibnamefont {Meyer}},\ }\href
  {http://www.pci.uni-heidelberg.de/tc/usr/mctdh/} {\emph {\bibinfo {title}
  {The {MCTDH} Package, Version 8.2}}} (\bibinfo {year} {2000})\BibitemShut
  {NoStop}%
\bibitem [{\citenamefont {Iung}\ \emph {et~al.}(2004)\citenamefont {Iung},
  \citenamefont {Gatti},\ and\ \citenamefont {Meyer}}]{MeyerHCF3}%
  \BibitemOpen
  \bibfield  {author} {\bibinfo {author} {\bibfnamefont {C.}~\bibnamefont
  {Iung}}, \bibinfo {author} {\bibfnamefont {F.}~\bibnamefont {Gatti}}, \ and\
  \bibinfo {author} {\bibfnamefont {H.-D.}\ \bibnamefont {Meyer}},\ }\href@noop
  {} {\bibfield  {journal} {\bibinfo  {journal} {J. Chem. Phys.}\ }\textbf
  {\bibinfo {volume} {120}},\ \bibinfo {pages} {6992} (\bibinfo {year}
  {2004})}\BibitemShut {NoStop}%
\bibitem [{\citenamefont {Gatti}\ and\ \citenamefont
  {Meyer}(2004)}]{Meyertolueno}%
  \BibitemOpen
  \bibfield  {author} {\bibinfo {author} {\bibfnamefont {F.}~\bibnamefont
  {Gatti}}\ and\ \bibinfo {author} {\bibfnamefont {H.-D.}\ \bibnamefont
  {Meyer}},\ }\href@noop {} {\bibfield  {journal} {\bibinfo  {journal} {Chem.
  Phys.}\ }\textbf {\bibinfo {volume} {304}},\ \bibinfo {pages} {3} (\bibinfo
  {year} {2004})}\BibitemShut {NoStop}%
\bibitem [{\citenamefont {Pasin}\ \emph {et~al.}(2006)\citenamefont {Pasin},
  \citenamefont {Gatti}, \citenamefont {Iung},\ and\ \citenamefont
  {Meyer}}]{MeyerHFCO}%
  \BibitemOpen
  \bibfield  {author} {\bibinfo {author} {\bibfnamefont {G.}~\bibnamefont
  {Pasin}}, \bibinfo {author} {\bibfnamefont {F.}~\bibnamefont {Gatti}},
  \bibinfo {author} {\bibfnamefont {C.}~\bibnamefont {Iung}}, \ and\ \bibinfo
  {author} {\bibfnamefont {H.-D.}\ \bibnamefont {Meyer}},\ }\href@noop {}
  {\bibfield  {journal} {\bibinfo  {journal} {J. Chem. Phys.}\ }\textbf
  {\bibinfo {volume} {124}},\ \bibinfo {eid} {194304} (\bibinfo {year}
  {2006})}\BibitemShut {NoStop}%
\bibitem [{\citenamefont {Pasin}\ \emph {et~al.}(2007)\citenamefont {Pasin},
  \citenamefont {Iung}, \citenamefont {Gatti},\ and\ \citenamefont
  {Meyer}}]{MeyerDFCO}%
  \BibitemOpen
  \bibfield  {author} {\bibinfo {author} {\bibfnamefont {G.}~\bibnamefont
  {Pasin}}, \bibinfo {author} {\bibfnamefont {C.}~\bibnamefont {Iung}},
  \bibinfo {author} {\bibfnamefont {F.}~\bibnamefont {Gatti}}, \ and\ \bibinfo
  {author} {\bibfnamefont {H.-D.}\ \bibnamefont {Meyer}},\ }\href@noop {}
  {\bibfield  {journal} {\bibinfo  {journal} {J. Chem. Phys.}\ }\textbf
  {\bibinfo {volume} {126}},\ \bibinfo {eid} {024302} (\bibinfo {year}
  {2007})}\BibitemShut {NoStop}%
\bibitem [{\citenamefont {Pasin}\ \emph {et~al.}(2008)\citenamefont {Pasin},
  \citenamefont {Iung}, \citenamefont {Gatti}, \citenamefont {Richter},
  \citenamefont {L\'eonard},\ and\ \citenamefont {Meyer}}]{HFCODFCO}%
  \BibitemOpen
  \bibfield  {author} {\bibinfo {author} {\bibfnamefont {G.}~\bibnamefont
  {Pasin}}, \bibinfo {author} {\bibfnamefont {C.}~\bibnamefont {Iung}},
  \bibinfo {author} {\bibfnamefont {F.}~\bibnamefont {Gatti}}, \bibinfo
  {author} {\bibfnamefont {F.}~\bibnamefont {Richter}}, \bibinfo {author}
  {\bibfnamefont {C.}~\bibnamefont {L\'eonard}}, \ and\ \bibinfo {author}
  {\bibfnamefont {H.-D.}\ \bibnamefont {Meyer}},\ }\href@noop {} {\bibfield
  {journal} {\bibinfo  {journal} {J. Chem. Phys.}\ }\textbf {\bibinfo {volume}
  {129}},\ \bibinfo {eid} {144304} (\bibinfo {year} {2008})}\BibitemShut
  {NoStop}%
\bibitem [{\citenamefont {Richerme}\ \emph {et~al.}(2014)\citenamefont
  {Richerme}, \citenamefont {Gong}, \citenamefont {Lee}, \citenamefont {Sengo},
  \citenamefont {Smith}, \citenamefont {Foss-Feig}, \citenamefont {Michalakis},
  \citenamefont {Gorshkov},\ and\ \citenamefont {Monroe}}]{Richerme}%
  \BibitemOpen
  \bibfield  {author} {\bibinfo {author} {\bibfnamefont {P.}~\bibnamefont
  {Richerme}}, \bibinfo {author} {\bibfnamefont {Z.-X.}\ \bibnamefont {Gong}},
  \bibinfo {author} {\bibfnamefont {A.}~\bibnamefont {Lee}}, \bibinfo {author}
  {\bibfnamefont {C.}~\bibnamefont {Sengo}}, \bibinfo {author} {\bibfnamefont
  {J.}~\bibnamefont {Smith}}, \bibinfo {author} {\bibfnamefont
  {M.}~\bibnamefont {Foss-Feig}}, \bibinfo {author} {\bibfnamefont
  {S.}~\bibnamefont {Michalakis}}, \bibinfo {author} {\bibfnamefont {A.~V.}\
  \bibnamefont {Gorshkov}}, \ and\ \bibinfo {author} {\bibfnamefont
  {C.}~\bibnamefont {Monroe}},\ }\href@noop {} {\bibfield  {journal} {\bibinfo
  {journal} {Nature}\ }\textbf {\bibinfo {volume} {511}},\ \bibinfo {pages}
  {198} (\bibinfo {year} {2014})}\BibitemShut {NoStop}%
\bibitem [{\citenamefont {Eisert}\ \emph {et~al.}(2015)\citenamefont {Eisert},
  \citenamefont {Friesdorf},\ and\ \citenamefont {Gogolin}}]{Eisert}%
  \BibitemOpen
  \bibfield  {author} {\bibinfo {author} {\bibfnamefont {J.}~\bibnamefont
  {Eisert}}, \bibinfo {author} {\bibfnamefont {M.}~\bibnamefont {Friesdorf}}, \
  and\ \bibinfo {author} {\bibfnamefont {C.}~\bibnamefont {Gogolin}},\
  }\href@noop {} {\bibfield  {journal} {\bibinfo  {journal} {Nat. Phys.}\
  }\textbf {\bibinfo {volume} {11}},\ \bibinfo {pages} {124} (\bibinfo {year}
  {2015})}\BibitemShut {NoStop}%
\bibitem [{\citenamefont
  {Suba\ifmmode\mbox{\c{s}}\else\c{s}\fi{}\ifmmode\imath\else\i\fi{}}\ \emph
  {et~al.}(2012)\citenamefont
  {Suba\ifmmode\mbox{\c{s}}\else\c{s}\fi{}\ifmmode\imath\else\i\fi{}},
  \citenamefont {Fleming}, \citenamefont {Taylor},\ and\ \citenamefont
  {Hu}}]{fleming}%
  \BibitemOpen
  \bibfield  {author} {\bibinfo {author} {\bibfnamefont {Y.}~\bibnamefont
  {Suba\ifmmode\mbox{\c{s}}\else\c{s}\fi{}\ifmmode\imath\else\i\fi{}}},
  \bibinfo {author} {\bibfnamefont {C.~H.}\ \bibnamefont {Fleming}}, \bibinfo
  {author} {\bibfnamefont {J.~M.}\ \bibnamefont {Taylor}}, \ and\ \bibinfo
  {author} {\bibfnamefont {B.~L.}\ \bibnamefont {Hu}},\ }\href@noop {}
  {\bibfield  {journal} {\bibinfo  {journal} {Phys. Rev. E}\ }\textbf {\bibinfo
  {volume} {86}},\ \bibinfo {pages} {061132} (\bibinfo {year}
  {2012})}\BibitemShut {NoStop}%
\bibitem [{\citenamefont {Heyl}(2017)}]{Heyl}%
  \BibitemOpen
  \bibfield  {author} {\bibinfo {author} {\bibfnamefont {M.}~\bibnamefont
  {Heyl}},\ }\href@noop {} {\bibfield  {journal} {\bibinfo  {journal}
  {arXiv:1709.07461v1 [cond-mat.stat-mech]}\ } (\bibinfo {year}
  {2017})}\BibitemShut {NoStop}%
\end{thebibliography}%

\end{document}